\documentclass[twocolumn,preprintnumbers,amsmath,amssymb,superscriptaddress,nofootinbib,longbibliography]{revtex4-1}
\pdfoutput=1

\usepackage{graphicx}
\usepackage{bm}
\usepackage{dsfont}
\usepackage[usenames,dvipsnames]{xcolor}
\usepackage{pstricks}
\usepackage[tight]{subfigure}
\usepackage{verbatim}
\usepackage{units}
\usepackage{multirow}
\usepackage{enumitem}
\usepackage{mathrsfs}
\usepackage{leftidx}
\usepackage{xspace}

\usepackage[a4paper]{hyperref}
\hypersetup{colorlinks=true,linktoc=all,linkcolor=blue,breaklinks=true,citecolor=blue,urlcolor=blue}

\usepackage[english]{babel}
\newcommand{\ket}[1]{|#1\rangle}

\newcommand{\GF}[2]{\langle\!\langle #1\vert#2\rangle\!\rangle}
\newcommand{\acG}{\mathcal{G}}
\newcommand{\sfA}{\mathcal{A}}

\newcommand{\TK}{T_\text{K}}
\newcommand{\TKqd}{T_\text{K}^0}

\newcommand{\GAP}{G^\text{AP}}
\newcommand{\GP}{G^\text{P}}

\newcommand{\SP}{\mathcal{P}}
\newcommand{\SPAP}{\mathcal{P}^\text{AP}}
\newcommand{\SPP}{\mathcal{P}^\text{P}}

\newcommand{\omegar}{\omega_\text{r}}

\newcommand{\Ham}{\skew{3}{\hat}{\mathcal{H}}}
\newcommand{\opS}{\skew{3}{\hat}{S}}
\newcommand{\ops}{\skew{2}{\hat}{s}}
\newcommand{\opc}{\skew{2}{\hat}{c}}
\newcommand{\opn}{\skew{2}{\hat}{n}}
\newcommand{\opa}{\skew{1}{\hat}{a}}

\newcommand{\opQ}{\skew{2}{\hat}{Q}}
\newcommand{\opI}{\skew{3}{\hat}{I}}
\newcommand{\opII}{\skew{3}{\hat}{\mathcal{I}}}
\newcommand{\oper}[1]{\skew{3}{\hat}{#1}}

\newcommand{\intd}{\text{d}}
\newcommand{\via}{\emph{via} }

\usepackage[normalem]{ulem}

\begin{document}


\title{Dynamical spin accumulation in large-spin magnetic molecules}

\author{Anna P\l omi\'nska}
\email{anna.plominska@amu.edu.pl}
\affiliation{Faculty of Physics, Adam Mickiewicz University, 61-614 Pozna\'{n}, Poland}
\author{Ireneusz Weymann}
\email{weymann@amu.edu.pl}
\affiliation{Faculty of Physics, Adam Mickiewicz University, 61-614 Pozna\'{n}, Poland}
\author{Maciej Misiorny}
\email{misiorny@amu.edu.pl}
\affiliation{Department of Microtechnology and Nanoscience MC2, Chalmers University of Technology, SE-412 96 G\"{o}teborg, Sweden}
\affiliation{Faculty of Physics, Adam Mickiewicz University, 61-614 Pozna\'{n}, Poland}

\date{\today}

\begin{abstract}
The frequency-dependent transport through a nano-device containing a large-spin magnetic molecule is studied theoretically in the Kondo regime.
Specifically, the effect of magnetic anisotropy on dynamical spin accumulation is of primary interest. Such accumulation arises due to finite off-diagonal in spin components of the dynamical conductance.
Here, employing the Kubo formalism and the numerical renormalization group (NRG) method, we demonstrate that the dynamical transport properties strongly depend on magnetic configuration of the device and intrinsic parameters of the molecule. 
Specifically, the effect of dynamical spin accumulation is found to be greatly affected by the type of  magnetic anisotropy exhibited by the molecule, and it develops for frequencies corresponding to the Kondo temperature.
For the parallel magnetic configuration of the device, the presence of dynamical spin accumulation is conditioned by the interplay of ferromagnetic-lead-induced exchange field and the Kondo correlations.
\end{abstract}

\maketitle

\section{Introduction}

Over the past two decades, nano-devices involving individual spin impurities strongly tunnel-coupled to leads have proven to be an excellent test-bed for studying quantum many-body effects in electronic transport, among which the Kondo effect is one of the most prominent ones~\cite{Scott_ACSNano4/2010,Hewson_book,Coleman_book}. In principle, the role of such a spin impurity can be played by any system that either inherently exhibits spin or is capable to accommodate a~single conduction electron, which has been experimentally demonstrated for various nanoscopic structures, such as, quantum dots~\cite{Goldhaber_Nature391/98,Cronenwett_Science281/1998,Sasaki_Nature405/2000,Nygaard_Nature(London)408/2000}, magnetic adatoms~\cite{Li_Phys.Rev.Lett.80/1998,Knorr_Phys.Rev.Lett.88/2002,Otte_NaturePhys.4/2008,Ternes_J.Phys.:Condens.Matter21/2009,Jacobson_Nat.Commun.6/2015} or molecules~\cite{Liang_Nature417/2002,Park_Nature417/2002,Kubatkin_Nature425/2003,Zhao_Science309/2005,Roch_Nature453/2008,Parks_Science328/2010,Zyazin_Synth.Met.161/2011,Choi_NanoLett.14/2014,Frisenda_NanoLett.15/2015,Liu_Phys.Rev.Lett.114/2015}.
A proper understanding of the effect of charge and spin correlations on electronic transport is especially sought for devices based on \emph{large-spin}~(\mbox{$S>1/2$}) impurities. Importantly, such systems are  a suitable platform for applications in emerging technologies for storage and processing information~\cite{Bogani_NatureMater.7/2008,Bartolome_book}, whose aim is to utilize magnetic properties of single atoms~\cite{Khajetoorians_Science352/2016,Donati_Science352/2016,Natterer_Nature543/2017} or molecules~\cite{Mannini_NatureMater.8/2009,Vincent_Nature488/2012,Cornia_NatureMater.16/2017,Goodwin_Nature548/2017}.
The key property of a large spin to serve as a base for a memory device is the uniaxial magnetic anisotropy that introduces an~energy barrier for spin reversal~\cite{Gatteschi_book}. The uniaxial component of magnetic anisotropy is, however, often accompanied by the transverse one~\cite{Misiorny_Phys.Rev.B91/2015} that, allowing for the~under-barrier transitions~\cite{Mannini_Nature468/2010}, has the parasitic effect on the~spin stability.
In the \emph{stationary} transport regime, the~interplay between the Kondo correlations and the magnetic anisotropy of a spin impurity has been predicted  to significantly  affect transport characteristics of a device. Specifically, this interplay leads to a number of spectroscopic features ranging from the current suppression due to the spin reversal barrier~\cite{Zitko_Phys.Rev.B78/2008,Zitko_NewJ.Phys.12/2010,Misiorny_Phys.Rev.Lett.106/2011,Misiorny_Phys.Rev.B84/2011} to some more intricate, Berry's phase-related effects originating from the quantum tunneling of spin~\cite{Romeike_Phys.Rev.Lett.96/2006,Romeike_Phys.Rev.Lett.97/2006E,Leuenberger_Phys.Rev.Lett.97/2006,Gonzalez_Phys.Rev.Lett.98/2007,Gonzalez_Phys.Rev.B78/2008,Misiorny_Phys.Rev.B90/2014,Romero_Phys.Rev.B90/2014}.

In the present paper, on the other hand, we address\linebreak the~\emph{dynamical} aspect of spin-dependent transport through
magnetic molecules in the Kondo regime.
Whereas this problem has been studied for spin one-half impurities~\cite{Sindel_Phys.Rev.Lett.94/2005,Toth_Phys.Rev.B76/2007,
Moca_Phys.Rev.B81/2010,Moca_Phys.Rev.B83/2011,Moca_Phys.Rev.B84/2011,Weymann_J.Appl.Phys.109/2011,Moca_Phys.Rev.B89/2014,Chirla_Phys.Rev.B89/2014,Hemingway_Phys.Rev.B90/2014,DiasdaSilva_Phys.Rev.B92/2015}, it has only recently attracted some attention in the context of large-spin impurities~\cite{Plominska_Phys.Rev.B.95/2017}. 
In~general, by analyzing the dynamical response of a~system to an~external time-dependent bias one obtains a~direct access to the fluctuations in the system, which is ensured by the fluctuation-dissipation theorem~\cite{Kubo_book} linking the dynamical conductance of the system with its~noise power spectral density.

Here, we specifically focus on the influence of magnetic anisotropy on the effect of \emph{dynamical spin accumulation}, which can be attributed to the non-zero off-diagonal in spin component of  frequency-dependent conductance~\cite{Moca_Phys.Rev.B81/2010,Moca_Phys.Rev.B84/2011}. The physical meaning of such accumulation can be better understood if one imagines that it actually corresponds to the situation when, for instance, one injects electrons of given spin orientation but detects the current of opposite spin direction.
For this purpose, we consider a magnetic molecule as an~exemplar of a large-spin impurity. Formed by a single conducting orbital exchange-coupled to an anisotropic magnetic core, such a model of a molecule captures effects both due to charging and magnetic anisotropy. 
The dynamical linear-response transport characteristics of the system are obtained using a combination of the Kubo approach~\cite{Kubo_book,Plominska_Phys.Rev.B.95/2017} and the numerical renormalization group~(NRG) method~\cite{Wilson_Rev.Mod.Phys.47/1975,Bulla_Rev.Mod.Phys.80/2008}.

We show that the effect of dynamical spin accumulation strongly depends on the intrinsic parameters of the~molecule and the magnetic configuration of the device.
For the antiparallel magnetic configuration and in the case of an easy-axis type of uniaxial magnetic anisotropy, the spin accumulation becomes generally suppressed, however, it can be restored if a transverse anisotropy component is also present.
This is contrary to the case of an easy-plane type of anisotropy, where a pronounced dynamical spin accumulation can develop both in the absence and presence of transverse anisotropy. 
For all considered cases, we find that a local maximum in the dynamical spin accumulation emerges for energy scale corresponding to the Kondo temperature, with the height dependent on the strength of Kondo correlations.
Furthermore, for the parallel magnetic configuration of the device, we demonstrate that the presence and magnitude of dynamical spin accumulation is conditioned by the interplay of ferromagnetic-lead proximity-induced quadrupolar exchange field and the correlations responsible for the formation of the Kondo effect.

The paper is organized as follows: In Sec.~\ref{sec:Model} an account of basic premises and assumptions of the model is provided, while in Sec.~\ref{sec:Dyn_cond} the theoretical framework used in calculations of the dynamical conductance is outlined. Numerical results are discussed in Sec.~\ref{sec:Results}, which  begins  with a detailed review of energy reference scales and model parameters~(Sec.~\ref{sec:Parameters}). The analysis of the~results we start for an \emph{anisotropic} molecule, and next, we include stepwise the \emph{uniaxial}~(Sec.~\ref{sec:Uniaxial}) and \emph{transverse}~(Sec.~\ref{sec:Uniaxial_and_transverse}) component of magnetic anisotropy into the picture. The effects of the spin polarization and magnetic configuration of electrodes  are discussed in Sec.~\ref{sec:Spin_polarization} and~Sec.~\ref{sec:Parallel}, respectively.
In Sec.~\ref{sec:AFM-J} the~transport behavior in the case of the antiferromagnetic coupling
between the molecule's core spin and the~orbital level is discussed.
Finally, the main conclusions are presented in Sec.~\ref{sec:Con}

\section{\label{sec:Model}Model system}

%
\begin{figure}[t]
	\includegraphics[scale=0.65]{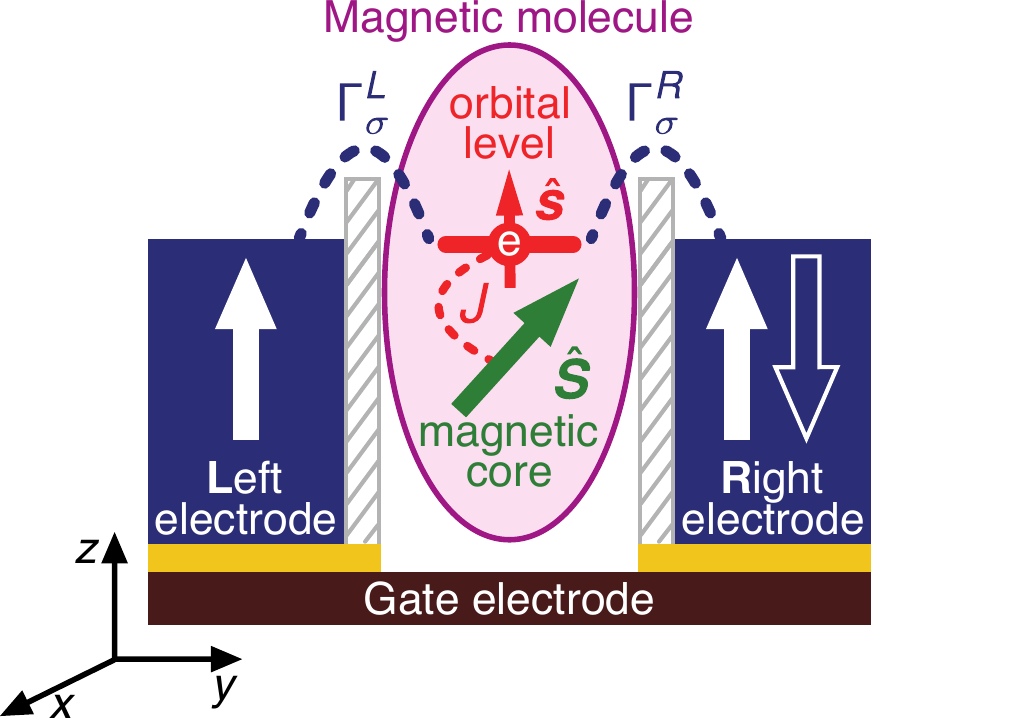}
	\caption{
    Schematic depiction of a large-spin magnetic mo\-le\-cule embedded in a magnetic tunnel junction. For a detailed description of the model system see Sec.~\ref{sec:Model}.
    \label{fig1}
    }
\end{figure}

In order to study the effect of dynamical spin accumulation in the case of a large-spin impurity, we employ here the model system that consists of a large-spin magnetic molecule embedded in the magnetic tunnel junction, see Fig.~\ref{fig1}.
Specifically, the magnetic molecule is represented as a large internal spin~$\oper{\bm{S}}$ (referred to also as magnetic core), with $S>1/2$, coupled \via exchange interaction~$J$ to a single orbital level (OL). It is assumed that the molecule is tunnel-coupled to ferromagnetic electrodes of the junctions only through the~OL, which essentially means that transport of electrons across the junction takes place exclusively through this orbital \cite{Elste_Phys.Rev.B73/2006,Misiorny_Phys.Rev.B79/2009}.
Moreover, spin-dependent electron tunneling processes between the OL and electrodes lead to broadening of the former, and this broadening is described by the spin-dependent hybridization function~$\Gamma^{q}_{\sigma}$ with $q = L(\text{eft}), R(\text{ight})$.

The full Hamiltonian~$\Ham_\textrm{total}$ characterizing the system under consideration has the following form
\begin{equation}\label{eq:H_total}
	\Ham_\textrm{total}
	=
	 \Ham_\textrm{OL}
	 + \Ham_\textrm{core}
	 + \Ham_\textrm{OL-core}
	 + \Ham_\textrm{el}
	 + \Ham_\textrm{tun}.
\end{equation}
Here, the first three terms are related to the magnetic molecule. In particular,~$\Ham_\textrm{OL}$ accounts for the key properties of the~con\-duct\-ing OL and it reads as
\begin{equation}\label{eq:H_OL}
	\Ham_\textrm{OL}
	=
	\varepsilon\sum_\sigma\opn_\sigma
	+
	U\opn_\uparrow \opn_\downarrow
	,
\end{equation}
with the first term representing the contribution due to occupation of the OL by an electron of spin~$\sigma$ and energy~$\varepsilon$, and the second term including the Coulomb interaction~$U$ that arises in the situation when two electrons of opposite spins reside in the OL. The relevant  occupation operator~$\opn_\sigma=\opc_\sigma^\dagger \opc_\sigma^{}$ is defined in terms of electron creation~($\opc_\sigma^\dagger$) and annihilation~$(\opc_\sigma^{})$ operators for the OL. We note that application of a voltage to the gate electrode allows for tuning the OL energy~$\varepsilon$.
Furthermore, the second term of Hamiltonian~(\ref{eq:H_total}) describes magnetic anisotropy of the molecule's magnetic core within the \emph{giant-spin approach}~\cite{Gatteschi_book},
\begin{equation}\label{eq:H_core}
	\Ham_\textrm{core}
	=
	-D\opS_z^2
	+
	E\big(\opS_x^2-\opS_y^2\big)
	,
\end{equation}
with $D$ and $E$ denoting the uniaxial and transverse magnetic anisotropy constants, respectively.
Finally, the exchange interaction between the magnetic core effective spin~$\oper{\bm{S}}$ and the spin of a single electron occupying the orbital
\mbox{
$
	\oper{\bm{s}} =
	(1/2)\sum_{\sigma\sigma^\prime}
	\hat{\bm{\sigma}}_{\sigma\sigma^\prime}
	\opc_\sigma^\dagger\opc_{\sigma^\prime}^{}
$,
}
where $\hat{\bm{\sigma}}\equiv(\hat{\sigma}_x,\hat{\sigma}_y,\hat{\sigma}_z)$ stands for the Pauli spin operator, is given by
\begin{equation}\label{eq:H_J}
	\Ham_\text{OL-core}
	=
	-J\oper{\bm{s}}\cdot\oper{\bm{S}},
\end{equation}
with the $J$-coupling being \emph{ferromagnetic} (FM) for $J>0$ and \emph{antiferromagnetic} (AFM) for $J<0$.
%

%
%
Ferromagnetic electrodes of the junction are approximated as reservoirs of non-interacting and spin-polarized electrons and described by the Hamiltonian
\begin{equation}\label{eq:H_el}
	\Ham_\text{el}
	=
	\sum_{q\sigma}
    \int_{-W}^W\intd\epsilon\,
	\epsilon\,
	\opa_\sigma^{q\dag}(\epsilon)
	\opa_\sigma^q(\epsilon)
	,
\end{equation}
where $\opa_\sigma^{q\dagger}(\epsilon)\ \big[\opa_\sigma^{q}(\epsilon)\big]$ is the operator responsible for creation [annihilation] of a spin-$\sigma$ electron in the $q$th electrode, and~$W$ denotes the conduction band half-width.
Moreover, only the case of a collinear relative orientation of the spin moments of electrodes, that is, the parallel (P) and antiparallel (AP) magnetic configuration, is considered. It is also assumed that the orientation of these spin moments is collinear with the principal axis of the molecule corresponding to the uniaxial component of its magnetic anisotropy.

Ultimately, the single electron tunneling processes between the OL and electrodes are captured by the last term of Hamiltonian~(\ref{eq:H_total}),
\begin{equation}\label{eq:H_tun}
	\Ham_\textrm{tun}
	=
	\sum_{q\sigma}
	\sqrt{\frac{\Gamma_\sigma^q}{\pi}}
    \int_{-W}^W\intd\epsilon\,
	\Big[
	\opa_\sigma^{q\dagger}(\epsilon)
	\opc_\sigma^{}
	+
	\opc_\sigma^\dagger
	\opa_\sigma^q (\epsilon)
	\Big]
	.
\end{equation}
Let us introduce the \emph{total broadening}
$
	\Gamma^q
	=
	\Gamma_\uparrow^q+\Gamma_\downarrow^q
$
of the OL due to its tunnel-coupling to the $q$th electrode, and define the \emph{spin polarization coefficient} for the $q$th electrode as
$
	p^q
	=
	\big(\Gamma_\uparrow^q-\Gamma_\downarrow^q\big)
	/
	\big(\Gamma_\uparrow^q+\Gamma_\downarrow^q\big)
$.
Next, assuming that both electrodes are made of the same material~($p^L=p^R\equiv p$) and that the OL is tunnel-coupled symmetrically to both electrodes~($\Gamma^L=\Gamma^R\equiv\Gamma$), one can parametrize the hybridization functions as follows:
$\Gamma^L_{\uparrow(\downarrow)}=\Gamma^R_{\uparrow(\downarrow)}=(\Gamma/2)(1\pm p)$ for the parallel magnetic configuration,
and $\Gamma^L_{\uparrow(\downarrow)}=\Gamma^R_{\downarrow(\uparrow)}=(\Gamma/2)(1\pm p)$ for the antiparallel one.

\section{\label{sec:Dyn_cond}
Dynamical system response}

Since the main goal is to analyze the effect of dynamical spin accumulation, below we outline a derivation of the frequency-dependent conductance (admittance) in terms of relevant spectral functions. In the next step, these functions will be calculated with the help of the Wilson's numerical renormalization group (NRG) method~\cite{Wilson_Rev.Mod.Phys.47/1975,Bulla_Rev.Mod.Phys.80/2008,Legeza_DMNRGmanual,Toth_Phys.Rev.B78/2008}.

To begin with, let us assume that an external bias voltage~$V^{L(R)}(t)$ modulated periodically in time is applied to the ferromagnetic electrodes. To take into account the effect of such a time-dependent bias, the full Hamiltonian~(\ref{eq:H_total}) of the system becomes modified by adding a new term~\cite{Toth_Phys.Rev.B76/2007,Moca_Phys.Rev.B81/2010,Weymann_J.Appl.Phys.109/2011,Plominska_Phys.Rev.B.95/2017},
\begin{equation}
	\Ham_\text{bias}
	=
	\sum_{q\sigma}
	\opQ_\sigma^q
	V^q(t)
	,
\end{equation}
with the operator~$\opQ_\sigma^q$ describing the spin-$\sigma$ component of charge induced in the $q$th electrode defined as
\begin{equation}\label{eq:charge_op}
	\opQ_\sigma^q
	=
	-|e|
	\int_{-W}^W\intd\epsilon\,
	\opa_\sigma^{q\dagger}(\epsilon)
	\opa_\sigma^{q}(\epsilon)
	.
\end{equation}

To calculate the current flowing through the system of a large-spin magnetic molecule,
we use the Kubo formula
\begin{equation}\label{eq:Kubo_formula}
	I^q(t)
	\equiv
	\langle \opI^q(t)\rangle
	=
	\sum_{q^\prime\sigma\sigma^\prime}
	\int\!\!\textrm{d}t^\prime\,
	\acG_{\sigma\sigma^\prime}^{qq^\prime}(t-t^\prime)
	V^{q^\prime}(t^\prime),
\end{equation}
where $\acG_{\sigma\sigma^\prime}^{qq^\prime}(t-t^\prime)$
stands for the time-dependent conductance and takes the following form
\begin{equation}\label{eq:G_vs_GF}
	\acG_{\sigma\sigma^\prime}^{qq^\prime}(t-t^\prime)
	=
	-\frac{i}{\hbar}\theta(t-t^\prime)
	\big\langle[
	\opI_\sigma^q(t),
	\opQ_{\sigma^\prime}^{q^\prime}(t^\prime)	
	]\big\rangle,
\end{equation}
with the current,~$\opI_\sigma^q(t)=\text{d}\opQ_\sigma^q(t)/\text{d}t$, and charge,~$\opQ_{\sigma^\prime}^{q^\prime}(t^\prime)$, operators given in the interaction picture, and~$\langle\ldots\rangle$ denoting the quantum-statistical average.
Next, after Fourier-transforming Eq.~(\ref{eq:G_vs_GF}) and performing laborious, albeit straightforward calculations, one can find a general expression for the frequency-dependent (dynamical) conductance~$\acG_{\sigma\sigma^\prime}^{qq^\prime}(\omega)$ ---for a detailed derivation see, e.g., Ref.~\cite{Plominska_Phys.Rev.B.95/2017}. 
At this point, let us focus on the current response~$I^R(\omega)$ in the right electrode, and use that this current is invariant under an overall potential shift by~\mbox{$-V^R(\omega)$}, which yields 
\begin{equation}\label{eq:IR_general}
	I^R(\omega)
	=
	\sum_{\sigma\sigma^\prime}
	\acG_{\sigma\sigma^\prime}^{RL}(\omega)
	\big[
	V^L(\omega)-V^R(\omega)
	\big]
	.
\end{equation}
Thus, taking into consideration only the real part of the \emph{right-left} component of the dynamical conductance,
$
	G^c(\omega)
	\equiv
	\sum_{\sigma\sigma^\prime}
	\big[
	\text{Re}\, \acG_{\sigma\sigma^\prime}^{RL}(\omega)
	\big]^c	
	,
$
for the parallel~($c=\text{P}$) and antiparallel~($c=\text{AP}$) magnetic configuration of the junction one obtains
\begin{equation}\label{eq:Gc_general}
	G^c(\omega)
	=	
	\sum_{\sigma\sigma^\prime}
	G_{\sigma\sigma^\prime}^c(\omega)
	,
\end{equation}
with the spin-resolved components~$G_{\sigma\sigma^\prime}^c(\omega)$ of the form
\begin{multline}\label{eq:Gc_general_spin}
	G_{\sigma\sigma^\prime}^c(\omega)
	=
	\frac{G_0}{2}
	\alpha_\sigma^c
	\bigg\{
	\delta_{\sigma\sigma^\prime}
	\big[\textrm{g}^{\text{OL}}_{\sigma}(\omega) ]^c
\\[-5pt]
    +
    \frac{1}{2}
	\beta_{\sigma\sigma^\prime}^c
	\big[\textrm{g}_{\sigma\sigma^\prime}^{\mathcal{I}}(\omega) \big]^c
	\bigg\}
	.
\end{multline}
The factors~$\alpha_\sigma^c$ and $\beta_{\sigma\sigma^\prime}^c$ in the equation above depend only on the magnetic configuration and the spin polarization coefficient~$p$ of electrodes,
\begin{equation}
	\alpha_\sigma^{\textrm{P}}
	=
	1+ \eta_\sigma p
	\quad
	\textrm{and}
	\quad
	\alpha_\sigma^{\textrm{AP}}
	=
	1- p^2,
\end{equation}
with $\eta_{\uparrow(\downarrow)}=\pm 1$,
\begin{equation}
	\beta_{\sigma\sigma^\prime}^\textrm{P}
	=
	\sqrt{\frac{1+\eta_{\sigma^\prime}p}{1+\eta_\sigma p}}
	\quad
	\textrm{and}
	\quad
	\beta_{\sigma\sigma^\prime}^\textrm{AP}
	=
	\frac{1+\eta_\sigma p}{1+\eta_{\sigma^\prime} p}
	.
\end{equation}
Furthermore, in Eq.~(\ref{eq:Gc_general_spin}),~$G_0\equiv2e^2/h$ is the conductance quantum, while~$\textrm{g}^{\text{OL}}_{\sigma}(\omega)$ and~$\textrm{g}_{\sigma\sigma^\prime}^{\mathcal{I}}(\omega)$ represent two different (dimensionless) contributions to the conductance~\cite{Moca_Phys.Rev.B81/2010},
\begin{equation}\label{eq:g_OL}
	\textrm{g}^{\text{OL}}_{\sigma}(\omega)
	=
	\frac{1}{2\omega}
	\int\!\intd\omega^\prime
	\,\!
	\frac{\sfA_\sigma^{\textrm{OL}}(\omega^\prime)}{\sfA_0}
    \big[
	f(\omega^\prime-\omega)-f(\omega^\prime+\omega)
	\big]
	,
\end{equation}
and
\begin{equation}\label{eq:g_I}
\begin{split}
	\textrm{g}^{ \mathcal{I}}_{\sigma\sigma^\prime}(\omega)
	=
	-
	\frac{1}{\omega}
	\cdot
	\frac{\sfA_{\sigma\sigma^\prime}^{\mathcal{I}}(\omega)}{\hbar\rho\sfA_0}.
\end{split}
\end{equation}
Importantly, the physical origin of each of these contributions can be deduced from analysis of the two spectral functions~$\sfA_\sigma^{\textrm{OL}}(\omega)$ and~$\sfA_{\sigma\sigma^\prime}^{\mathcal{I}}(\omega)$ occurring in Eqs.~(\ref{eq:g_OL}) and~(\ref{eq:g_I}), respectively.
In particular, the former spectral function describes the orbital level (OL),
and it is defined~as
\begin{equation}\label{eq:sfA_OL}
	\sfA_\sigma^{\textrm{OL}}(\omega)
	\equiv
	-
	\frac{1}{\pi}\,
	\textrm{Im}\GF{c_\sigma^{}}{c_\sigma^\dagger}_\omega^\textrm{r}
	.
\end{equation}
The latter, on the other hand, is given by
\begin{equation}\label{eq:sfA_I}
	A_{\sigma\sigma^\prime}^{\mathcal{I}}(\omega)
	\equiv
	-
	\frac{1}{\pi}\,
	\textrm{Im}\GF{\opII_\sigma}{\opII_{\sigma^\prime}^{\dagger}}_\omega^\textrm{r},
\end{equation}
and this spectral function is associated with the dimensionless current operator
\begin{equation}
	\opII_\sigma
	\equiv
	\opc_\sigma^\dagger \hat{\Psi}_\sigma^{}
	-
	\hat{\Psi}_\sigma^{\dagger}\opc_\sigma^{}
	.
\end{equation}
Here, the field operator~$\hat{\Psi}_\sigma^{}$ corresponds essentially to the even linear combination of electrode operators,
\begin{equation}
	\hat{\Psi}_\sigma
	=
	\sqrt{\rho}\int\intd\epsilon\,
	\Big[
	\Lambda_\sigma^L
	\opa_\sigma^L(\epsilon)
	+
	\Lambda_\sigma^R
	\opa_\sigma^R(\epsilon)
	\Big]
\end{equation}
with
$
	\Lambda_\sigma^q
	=
	\sqrt{\Gamma_\sigma^q/(\Gamma_\sigma^L + \Gamma_\sigma^R)}
$
and $\rho=1/(2W)$ being the density of states of a conduction band.
Finally, the scaling factor $\sfA_0=1/(\pi\Gamma)$ in Eqs.~(\ref{eq:g_OL})-(\ref{eq:g_I}) denotes the spectral function of a single-level quantum dot (or in other words, that of the OL disconnected from the internal spin, $J=0$) at $\omega=0$ and for nonmagnetic electrodes. The function~$f(\omega)$ in Eq.~(\ref{eq:g_OL}) is the Fermi-Dirac distribution,
$
	f(\omega)
	=
	\big\{
	1
	+
	\exp[\hbar \omega / k_\text{B} T] \big\}^{-1}
$,
with $T$ being temperature and $k_\text{B}$ standing for the Boltzmann constant.

At this point, we would like to emphasize that one of the main quantities of interest in this paper is associated with the off-diagonal in spin  components of the spin-resolved dynamical conductance. In particular, $G_{\uparrow\downarrow}^c(\omega)$ and~$G_{\downarrow\uparrow}^c(\omega)$ take into account the spin correlations between the spin-up and spin-down channels and their finite values can be associated with the effect of dynamical spin accumulation that can build up in the molecule at finite driving frequencies $\omega$ \cite{Moca_Phys.Rev.B81/2010,Moca_Phys.Rev.B84/2011}.

%
As one can see from Eqs.~(\ref{eq:Gc_general_spin}) and~(\ref{eq:g_OL})-(\ref{eq:g_I}), in order to calculate the spin-resolved components~$G_{\sigma\sigma^\prime}^c(\omega)$ of the dynamical conductance, one needs to know first the spectral functions: $\sfA_\sigma^{\textrm{OL}}(\omega)$, Eq.~(\ref{eq:sfA_OL}), and $A_{\sigma\sigma^\prime}^{\mathcal{I}}(\omega)$, Eq.~(\ref{eq:sfA_I}). In the present work, these functions are derived using the NRG method~\cite{Wilson_Rev.Mod.Phys.47/1975,Bulla_Rev.Mod.Phys.80/2008,Legeza_DMNRGmanual,Toth_Phys.Rev.B78/2008}.
The idea of NRG is based on the logarithmic discretization of the conduction band with the discretization parameter $\Lambda$. In the next step, such a discretized model is mapped onto a semi-infinite chain with exponentially decaying hoppings. Importantly, the first site of semi-infinite chain is coupled to a spin impurity.
To obtain the following results, we used the discretization parameter~\mbox{$\Lambda = 2$}, and we kept~\mbox{$N_k=2560$} states during calculations. The high accuracy of calculations was achieved by averaging
the spectral data over~\mbox{$N_z = 4$} different discretization meshes~\cite{OliveiraPhysRevB.49.11986}.
Moreover, when discussing the behavior of the conductance in the zero-frequency limit,
we will present the data obtained from the full density-matrix NRG approach
\cite{Legeza_DMNRGmanual,Weichselbaum_Phys.Rev.Lett.99/2007}
by assuming~\mbox{$T/W=10^{-12}$}, which is much smaller than
the other energy scales considered throughout this paper.

\section{\label{sec:Results}Numerical results and discussion}
%
The main goal of this paper is to investigate the effect of dynamical spin accumulation in large-spin magnetic molecules, and in particular, to discuss how this effect is affected when the molecular spin is subject to magnetic anisotropy.
For this purpose, we consider the model of a magnetic molecule introduced in Sec.~\ref{sec:Model} with the magnetic core characterized by spin~\mbox{$S=2$}. To conduct a systematic analysis of the problem, and to understand how the dynamical spin accumulation manifests itself in frequency-dependent transport characteristics, first we will address the simplest example of a \emph{spin-isotropic} molecule (\mbox{$D=0$} and~\mbox{$E=0$}). Next,  we will discuss in Sec.~\ref{sec:Uniaxial} how dynamical spin accumulation changes when the \emph{uniaxial} component of magnetic anisotropy becomes gradually involved (\mbox{$D\neq0$} and~\mbox{$E=0$}). Finally, also the \emph{transverse} component (\mbox{$D\neq0$} and~\mbox{$E\neq0$}) will be included in Sec.~\ref{sec:Uniaxial_and_transverse}  to establish the complete picture of the problem.

\subsection{\label{sec:Parameters}Energy reference scales and model parameters}

In the regime of strong tunnel coupling of a molecule to electrodes, which is of key interest here, transport properties of the system are determined by strong charge and spin correlations. As a result, one can generally expect the Kondo effect to play a dominant role as soon as the~OL is occupied by a single electron and temperature is lower than some characteristic energy scale, referred to as the Kondo temperature~$\TK$~\cite{Hewson_book}.
In general,~$\TK$ can depend on different parameters of a system under consideration. For instance, in the case of a single-level quantum dot (the Anderson impurity) attached to ferromagnetic electrodes $\TK$, in the vicinity of the particle-hole symmetry point (\mbox{$\varepsilon/U\approx-0.5$}), is determined by the Coulomb interaction~$U$, the broadening  of the level~$\Gamma$ and the spin polarization of electrodes~$p$~\cite{Haldane_Phys.Rev.Lett.40/1978,Martinek_Phys.Rev.Lett.91/2003_127203}. On the other hand, in large-spin systems the Kondo temperature can be additionally affected by other parameters of the model, such as, the spin length~$S$ or magnetic anisotropy constants~$D$ and~$E$~\cite{Romeike_Phys.Rev.Lett.96/2006,Romeike_Phys.Rev.Lett.97/2006E,Zitko_Phys.Rev.B78/2008,Misiorny_Phys.Rev.B90/2014}.
Thus, in the following calculations we use the Kondo temperature~$\TK$ for a bare~OL~(\mbox{$J=0$})
tunnel-coupled to nonmagnetic electrodes (\mbox{$p=0$}),
given in energy units (\mbox{$k_\text{B}\equiv1$}), as a consistent energy reference scale insensitive to magnetic properties of the molecule. Such a choice of~$\TK$ corresponds in fact to the Kondo temperature of a single-level quantum dot, and henceforth we will refer to this temperature as~$\TKqd$.
Specifically, the temperature dependence of zero-frequency normalized linear conductance~$\mbox{$G(\omega =0,T)$}/\mbox{$G(\omega =0,T=0)$}$ at the particle-hole symmetry point (\mbox{$\varepsilon/U=-0.5$}) is used for estimation of~$\TKqd$ from the following condition~$G\mbox{$(\omega =0,T=\TKqd)$}/\mbox{$G(\omega =0,T=0)$}=1/2$.

In this paper, all the results are obtained for $T=0$. With regard to the magnetic configuration of the junction, the main discussion is carried out for the case of the antiparallel orientation of spin moments in electrodes. In such a configuration the effective spintronic dipolar~\cite{Martinek_Phys.Rev.Lett.91/2003_127203} and quadrupolar~\cite{Misiorny_NaturePhys.9/2013} exchange fields are generally absent, which allows us to analyze how the dynamical spin accumulation is affected exclusively by the intrinsic molecular magnetic anisotropy. Later on, we will also include the spintronic contribution to magnetic anisotropy by switching the junction into the parallel magnetic configuration. Note that throughout the paper a molecule is assumed to be electrostatically tuned \via a gate electrode to the particle-hole symmetric point \mbox{$\varepsilon/U = -0.5$}, so that in the parallel magnetic configuration the dipolar exchange field does not arise \cite{Misiorny_Phys.Rev.Lett.106/2011}.
In this way, we can consistently exclude any effects stemming from the presence of a magnetic field,
either real or effective, which are not the subject of the present analysis.

Moreover, the Coulomb energy is chosen \mbox{$U/W=0.4$}, with the half-width of conduction band~$W$ serving here as the energy unit (that is, \mbox{$W\equiv 1$}),  whereas the magnitude of the constant~$J$ describing the exchange coupling between the OL and the electrodes is taken to be $|J|/W=0.0045$. Depending on whether~$J$ is \emph{positive} or \emph{negative}, one can expect either the \emph{underscreened}~\cite{Coleman_Phys.Rev.B68/2003,Posazhennikova_Phys.Rev.Lett.94/2005,Koller_Phys.Rev.B72/2005,Misiorny_Phys.Rev.B86/2012_UK} or \emph{two-stage}~\cite{Posazhennikova_Phys.Rev.B75/2007,Zitko_J.Phys.:Condens.Matter22/2010,Wojcik_PhysRevB.91.134422/2014} Kondo effect, respectively, to arise in the system~\cite{Misiorny_Phys.Rev.B86/2012}. Here, we focus on the case of~\mbox{$J>0$}, that is, on the ferromagnetic type of the $J$-coupling, and the key differences occurring for the antiferromagnetic coupling ($J<0$) will be addressed only at the end, in Sec.~\ref{sec:AFM-J}.
Finally, the broadening of the OL due to tunneling of electrons to/from external electrodes is assumed to be $\Gamma/U =0.1$, while the spin polarization of electrodes is $p=0.5$, unless stated otherwise. Consequently, for the parameters assumed above one finds the 
reference Kondo temperature to be $\TKqd/W=0.002$.

\subsection{The effect of magnetic anisotropy on the dynamical spin accumulation}
%
%
\begin{figure}[t]
	\includegraphics[scale=1.0]{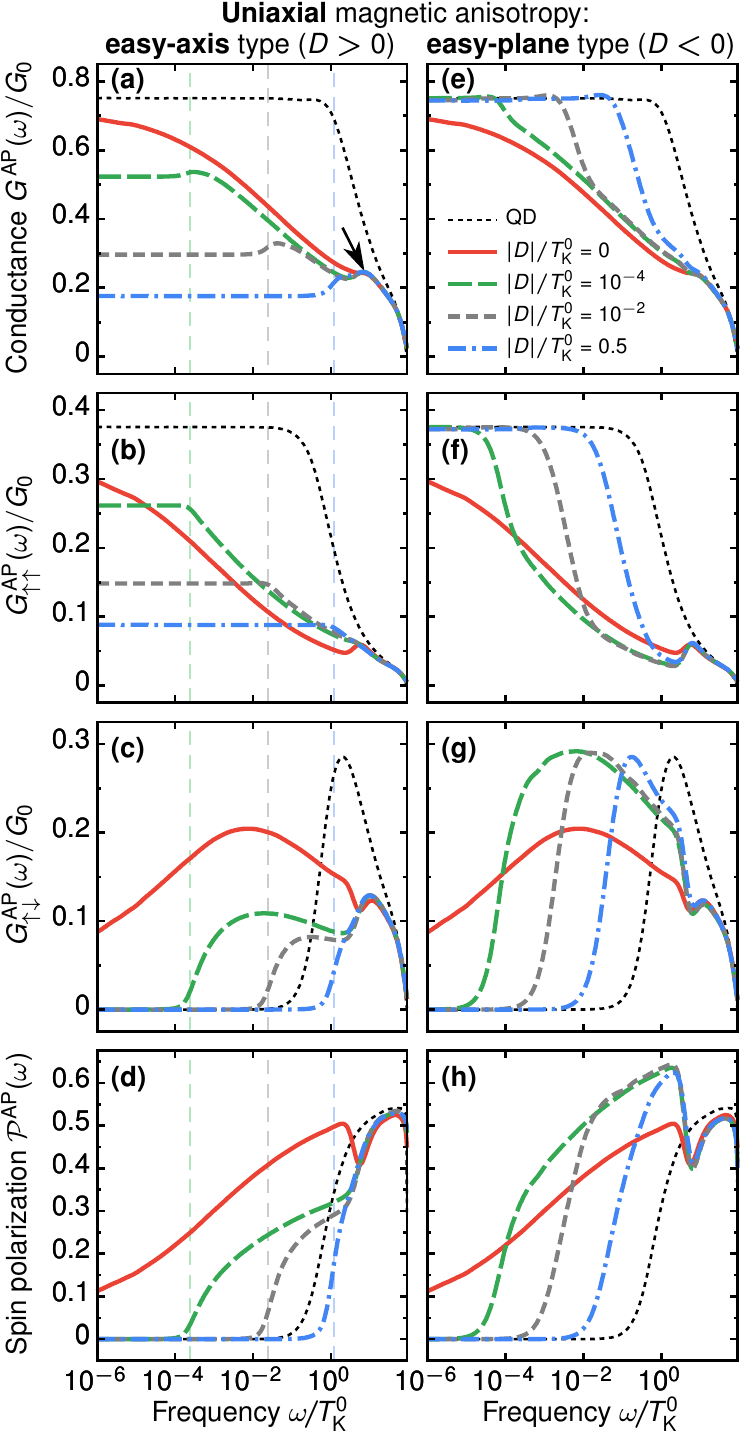}
	\caption{
	The effect of the \emph{uniaxial} component of mag\-ne\-tic anisotropy (\mbox{$D\neq0$}  and \mbox{$E=0$}) on the fre\-quency-dependent conductance of a large-spin magnetic molecule shown for the FM $J$-coupling and the junction in the antiparallel (AP) magnetic configuration.
   	\emph{Left} [\emph{right}] \emph{column} corresponds to the molecule exhibiting the \emph{easy-axis}~(\mbox{$D>0$}) [\emph{easy-plane}~(\mbox{$D<0$})] type of magnetic anisotropy.
   	\emph{Bottom panels}~(d,h) present the current spin polarization $\SPAP(\omega)$ which arises here solely due to the dynamical spin accumulation, i.e., \mbox{$\SPAP(\omega)=\SPAP_\text{dsa}(\omega)$}, Eq.~(\ref{eq:SP_dsa}). 
    The solid lines represent the case of a \emph{spin-isotropic} molecule, while the thin dotted lines are for a single-level quantum dot (QD), i.e., for~\mbox{$J=0$}. 
    The vertical dashed lines indicate the corresponding excitation energies,
    for details see the main text.
    The scaling factor~$G_0$ stands for the conductance quantum.
    For a discussion of parameters assumed in calculations see Sec.~\ref{sec:Parameters}.
    \label{fig2}
    }
	\vspace*{-20pt}
\end{figure}

Before we proceed to a discussion of how the dynamical spin accumulation is affected by the presence of magnetic anisotropy, it may be instructive to focus first briefly on frequency-dependence transport features of a \emph{spin-isotropic} molecule%
\footnote{%
Note that in order to enable qualitative comparison of the present results with the previous studies for a single-level quantum dot (QD)~\cite{Toth_Phys.Rev.B76/2007,Moca_Phys.Rev.B84/2011,Weymann_J.Appl.Phys.109/2011}, in Fig.~\ref{fig2} dotted lines representing the latter case have been added. However, in the main text we do not discuss this case. 
}.
This case is illustrated with the solid line in Fig.~\ref{fig2}, where the dynamical conductance~$\GAP(\omega)$ for the antiparallel magnetic configuration of the junction is shown as a function of frequency~$\omega$. Since~$\GAP(\omega)$ can be in general resolved into four spin components~$G_{\sigma\sigma^\prime}^\text{AP}(\omega)$, see Eq.~(\ref{eq:Gc_general}), in Fig.~\ref{fig2} apart from the total conductance~$\GAP(\omega)$~[plotted in (a,e)] we also present the spin-diagonal~$\GAP_{\sigma\sigma}(\omega)$~[in~(b,f)] and off-diagonal~$\GAP_{\sigma\overline{\sigma}}(\omega)$~[in~(c,g)] contributions, with the notation~$\overline{\sigma}$ to be read as \mbox{$\overline{\uparrow}\equiv\ \downarrow$} and \mbox{$\overline{\downarrow}\equiv\ \uparrow$}. In particular, here only the components for \mbox{$\sigma=\ \uparrow$} are shown, because in the antiparallel magnetic configuration the following symmetries hold:%
\footnote{The origin of the two symmetries can be explained by noting that the tunnel-coupling of a molecule to two electrodes, Eq.~(\ref{eq:H_tun}), can be effectively reduced to a single channel problem by applying an appropriate unitary transformation~\cite{Glazman_JETP.Lett.47/1988}. Interestingly, in the case of \emph{antiparallel} magnetic configuration of the junction the new effective spin-dependent tunnel coupling becomes actually independent of the spin polarization~$p$ of electrodes. As a result, calculations of the spectral functions~$\sfA_\sigma^\text{OL}(\omega)$, Eq.~(\ref{eq:sfA_OL}), and $\sfA_{\sigma\sigma^\prime}^{\mathcal{I}}(\omega)$, Eq.~(\ref{eq:sfA_I}), proceed as if electrodes were \emph{nonmagnetic}, so that eventually the values of these functions do not depend on spin indices.
}
\begin{equation*}
	\GAP_{\uparrow\uparrow}(\omega)
	\!=\!
	\GAP_{\downarrow\downarrow}(\omega)
 \;\;\;
{\rm and}  \;\;\;
	\GAP_{\downarrow \uparrow}(\omega)
	\!=\!
	\frac{(1\!-\!p)^2}{(1\!+\!p)^2}
	\GAP_{\uparrow \downarrow}(\omega).
\end{equation*}
Moreover, since the dynamical spin accumulation essentially leads to enhancement of imbalance in the number of electrons with opposite spin orientations transferred across the junction \via a molecule, we introduce here the frequency-dependent parameter~$\SP(\omega)$ characterizing the spin polarization of the current injected into a drain electrode defined as:
\begin{equation}
	\SP(\omega)
	\equiv 
	\frac{
	I_\uparrow(\omega)-I_\downarrow(\omega)
	}{
	I_\uparrow(\omega)+I_\downarrow(\omega)
	}
	.
\end{equation}
In the situation under consideration, the role of a drain is played by the right electrode, \mbox{$I_\sigma^{}(\omega)\equiv I_\sigma^R(\omega)$}, Eq.~(\ref{eq:IR_general}), and 
\mbox{$
	I_\sigma(\omega)
	\propto
	G_{\sigma\sigma}(\omega)
	+
	G_{\sigma\overline{\sigma}}(\omega)
$},
so that 
\begin{equation}\label{eq:SP}
	\SP(\omega)
	=
	\SP_0(\omega)
	+
	\SP_\text{dsa}(\omega)
	.
\end{equation}
Importantly, the current spin polarization parameter $\SP(\omega)$ consists of two terms: the first representing the diagonal in spin contribution to the conductance, 
\begin{equation}\label{eq:SP_0}
	\SP_0(\omega)
	=
	\frac{
	G_{\uparrow\uparrow}(\omega)-G_{\downarrow\downarrow}(\omega)
	}{
	\sum_{\sigma\sigma^\prime}G_{\sigma\sigma^\prime}(\omega)
	}	
	,
\end{equation}
and the second arising exclusively due to the dynamical spin accumulation, 
\begin{equation}\label{eq:SP_dsa}
	\SP_\text{dsa}(\omega)
	=
	\frac{
	G_{\uparrow\downarrow}(\omega)-G_{\downarrow\uparrow}(\omega)
	}{
	\sum_{\sigma\sigma^\prime}G_{\sigma\sigma^\prime}(\omega)
	}
	,
\end{equation}
which vanishes in the limit of~\mbox{$\omega\rightarrow0$} \cite{Moca_Phys.Rev.B84/2011}. Recall that the spin-resolved components $G_{\sigma\sigma^\prime}(\omega)$ of conductance are given by Eq.~(\ref{eq:Gc_general_spin}).  
One can, thus, immediately conclude that in the antiparallel magnetic configuration \mbox{$\SPAP_0(\omega)=0$}, and consequently, no spin polarization of the current occurs in the stationary case, \mbox{$\SPAP(\omega=0)=0$}, whereas for finite-frequency transport $\SPAP(\omega)=\SPAP_\text{dsa}(\omega)$ \mbox{---namely,} the spin polarization of the current is here a purely dynamical effect.

First of all, we recall that the conductance of a spin-isotropic system in the zero-frequency limit approaches \mbox{$\lim_{\omega\rightarrow0}\GAP(\omega)=(1-p^2)G_0$}, and~\mbox{$\GAP(\omega=0)$} consists solely of the diagonal-in-spin components, that is, \mbox{$\GAP(\omega=0)=\sum_\sigma\GAP_{\sigma\sigma}(\omega=0)$}, with the off-diagonal components being identically equal to zero, \mbox{$\GAP_{\sigma\overline{\sigma}}(\omega=0)=0$}. 
As the driving frequen\-cy~$\omega$ gets larger, one observes a monotonic decrease in~$\GAP(\omega)$, see Fig.~\ref{fig2}(a). However, a closer examination of spin components of the conductance reveals that whereas $\GAP_{\uparrow\uparrow}(\omega)$ decreases in value as well,
$\GAP_{\uparrow\downarrow}(\omega)$ actually follows the opposite trend and exhibit a maximum. The broad maximum in~$\GAP_{\uparrow\downarrow}(\omega)$ appears approximately at the frequency corresponding to the Kondo temperature~$\TK$~\cite{Plominska_Phys.Rev.B.95/2017}. As one may notice, here \mbox{$\TK\ll\TKqd$}, which stems from the fact that~$\TK$ becomes suppressed with the increase of~$J$~\cite{Misiorny_Phys.Rev.B86/2012}. Such a finite-frequency feature
in the off-diagonal in spin components of $\GAP(\omega)$ is a hallmark of the dynamical spin accumulation occurring in the system.
Moreover, unlike $\GAP_{\uparrow\downarrow}(\omega)$, the spin polarization~$\SPAP(\omega)$ of the current increases monotonically with~$\omega$ until \mbox{$\omega\approx\TKqd$}, where a local maximum occurs, see Fig.~\ref{fig2}(d).

On the other hand, in the limit of large frequencies,~\mbox{$\omega\gtrsim\TKqd$}, one can already point out that the effect of magnetic anisotropy is expected to be negligible (if~\mbox{$|D|<\TKqd$}). This stems from the fact that by further raising frequency~$\omega$, one in fact increases the amount of energy pumped into the system. As a result, excitations between states belonging to the two spin multiplets~\mbox{$S+1/2$} and~\mbox{$S-1/2$}, arising due to the $J$-coupling, take place, and they are observed as a small resonance in~$\GAP(\omega)$ at frequency~\mbox{$\omega\approx J$}, indicated by the arrow in Fig.~\ref{fig2}(a). Interestingly, such resonant transitions enhance the diagonal-in-spin conductance,
Fig.~\ref{fig2}(b), whereas they have a detrimental effect on the dynamical spin accumulation, leading to a local minimum in~$\GAP_{\uparrow\downarrow}(\omega)$, Fig.~\ref{fig2}(c), and consequently, also in the current spin polarization~$\SPAP(\omega)$, Fig.~\ref{fig2}(d).
Finally, with the further increase of~$\omega$ one expects that excitations related to the Coulomb interaction should become the key factor determining the behavior of conductance~\cite{Weymann_J.Appl.Phys.109/2011,Plominska_Phys.Rev.B.95/2017}. Since the purpose of the present work is to study the effects due to magnetic anisotropy, which take place at frequencies  corresponding to energy scales set by magnetic anisotropy constants~$D$ and~$E$, Eq.~(\ref{eq:H_core}), such a large-frequency regime (\mbox{$\omega\approx U$}) is not relevant and, thus, it will not be considered here.

\subsubsection{\label{sec:Uniaxial}Uniaxial magnetic anisotropy}

The situation changes when the molecular spin starts energetically preferring some specific spatial orientation(s), which essentially means that the spin is subject to magnetic anisotropy described generally by the Hamiltonian~(\ref{eq:H_core}). Let us first analyze the case when such a preferable orientation of the spin is related to a specific axis, customarily associated with the $z$ quantization axis, see Fig.~\ref{fig1}, and represented by the term~$-D\opS_z^2$ in Eq.~(\ref{eq:H_core}). One can immediately see that depending on the sign of the uniaxial magnetic anisotropy constant~$D$, the energy of the system becomes minimized if the molecular spin is oriented either along (for~\mbox{$D>0$}) or perpendicular to (for~\mbox{$D<0$}) the~$z$~axis. The former case is often referred to as the \emph{`easy-axis'} type of magnetic anisotropy, while the latter one as the \emph{`easy-plane'} type.
For systems characterized by a half-integer large spin (\mbox{$S>1/2$}), as the one considered here, it is a known, experimentally observed fact that their Kondo-dominated zero-frequency linear current response remains unaffected by the uniaxial magnetic anisotropy if~\mbox{$D<0$}~\cite{Otte_NaturePhys.4/2008,Ternes_J.Phys.:Condens.Matter21/2009,Ternes_NewJ.Phys.17/2015,Khajetoorians_NatureNanotechnol.10/2015}, whereas for~\mbox{$D>0$} the transport becomes suppressed~\cite{Parks_Phys.Rev.Lett.99/2007,Parks_Science328/2010,Zyazin_Synth.Met.161/2011,Liu_Phys.Rev.Lett.114/2015}.

The dynamical conductance for the case of uniaxial magnetic anisotropy of the easy-axis type~(\mbox{$D>0$}) is presented in the left column of Fig.~\ref{fig2}. Several distinctive features in~$\GAP(\omega)$ that evolve with the increase of~$D$ can be immediately spotted in Fig.~\ref{fig2}(a). To begin with, the zero-frequency conductance~\mbox{$\GAP(\omega=0)$} becomes reduced and $\GAP(\omega)$ remains frequency-independent as~$\omega$ grows. Once the frequency reaches \mbox{$\omega\approx D$}, a small peak forms, and for even larger values of~$\omega$ the conductance gets diminished.
Comparing the diagonal-in-spin component~$\GAP_{\uparrow\uparrow}(\omega)$ in Fig.~\ref{fig2}(b) with the off-diagonal one~$\GAP_{\uparrow\downarrow}(\omega)$ in Fig.~\ref{fig2}(c), one can conclude that the enhancement of the conductance at~\mbox{$\omega\approx D$} can be fully attributed to the effect of the dynamical spin accumulation. Importantly, it can be noticed that, unlike in the spin-isotropic case where the spin accumulation, \mbox{$\SPAP(\omega)\neq0$}, persists over a wide range of frequencies, now the effect arises only above some threshold frequency~$\omega^\ast$. Moreover, for \mbox{$\omega\gtrsim\omega^\ast$} the diagonal component~\mbox{$\GAP_{\uparrow\uparrow}(\omega)$} decreases monotonically (until \mbox{$\omega\approx\TKqd$}), whereas \mbox{$\GAP_{\uparrow\downarrow}(\omega)$} first builds up rapidly and then changes only insignificantly up to the limit of $\omega\gtrsim\TKqd$. As a result, the current spin polarization~$\SPAP(\omega)$, similarly as in the spin-isotropic case, increases steadily for larger and lager~$\omega$, though the achievable values of~$\SPAP(\omega)$ are appreciably smaller than for a spin-isotropic molecule.

The presence of the threshold frequency~$\omega^\ast$ above which the dynamical spin accumulation takes place can be understood by considering the mechanism underlying the spin-exchange (Kondo) processes responsible for flipping the spin orientation of an electron in the molecular OL. To gain an intuitive picture of such processes, it is instructive to analyze the eigenstates of a free-standing molecule, described by the Hamiltonian
\mbox{
$
	\Ham_\text{mol}
	=
	\Ham_\text{OL}
	+
	\Ham_\text{core}
	+
	\Ham_\text{OL-core}
	,
$
}
see Eqs.~(\ref{eq:H_OL})-(\ref{eq:H_J}), which participate in transport. Since the results are obtained at~\mbox{$T=0$}, it suffices to consider only the states of lowest energy.
For~\mbox{$J>0$}, the~ground state doublet of the $S+1/2$ spin multiplet has the form
\mbox{
$
	\ket{S_z^\text{tot}=\pm5/2}
	\equiv
	\ket{\pm1/2}_\text{OL}
	\otimes
	\ket{\pm2}_\text{core}
	,
$
}
with $\ket{s_z(S_z)}_{\text{OL(core)}}$ denoting the spin state of the~OL (magnetic core). Because the spin-exchange processes, occurring in the OL due to its strong hybridization to electrodes, can just lead to flipping of the OL spin, \mbox{$\ket{-1/2}_\text{OL}$}$\,\leftrightarrow\,$\mbox{$\ket{1/2}_\text{OL}$}, without affecting the state of the internal spin~$\ket{S_z}_\text{core}$, no direct transitions between the doublet ground states are possible.
In fact, any pair of molecular states~$\ket{S_{z,1}^\text{tot}}$ and~$\ket{S_{z,2}^\text{tot}}$ can support the spin-exchange processes due to tunneling of electrons only if~$|S_{z,1}^\text{tot}-S_{z,2}^\text{tot}|=1$, which basically stems from the fact that angular momentum exchanged between the tunneling current and the molecule has to be conserved.
It means that the only allowed transitions from the states~\mbox{$\ket{S_z^\text{tot}=-5/2}$} and~\mbox{$\ket{S_z^\text{tot}=5/2}$} can be those to the first excited doublet
states%
\footnote{Note that the states~\mbox{$\ket{S_z^\text{tot}=\pm3/2}$} constituting the first excited doublet of the~\mbox{$S+1/2$} spin multiplet have the from of a linear combination of the following states:
\mbox{
$
	\big\{
	\ket{\pm1/2}_\text{OL}
	\otimes
	\ket{\pm1}_\text{core}
	,
$}
\mbox{$
	\ket{\mp1/2}_\text{OL}
	\otimes
	\ket{\pm2}_\text{core}	
	\big\}
	.
$}
}
\mbox{$\ket{S_z^\text{tot}=-3/2}$} and~\mbox{$\ket{S_z^\text{tot}=3/2}$}, respectively. Importantly, these states are separated from the ground state doublet by an energy gap~$\Delta$. As a result, the spin exchange processes, which underlie the dynamical spin accumulation, become active if the energy pumped into the molecule by means of periodic driving external potential satisfies the condition~\mbox{$\omega\gtrsim\omega_{D>0}^\ast\approx\Delta$}. In the limit of \mbox{$D\ll J$} and at the particle-hole symmetry point ($\varepsilon/U=-0.5$), one can thus estimate~\cite{Misiorny_Phys.Rev.B84/2011}:~\mbox{$\omega_{D>0}^\ast\approx K_SD$} with
\mbox{
$
	K_S
	=
	2
	S(2S-1)/(2S+1)
$.
}
Those excitation energies are marked in left column of Fig.~\ref{fig2} with vertical dashed lines.
As can be seen, the agreement between this estimate and numerical data is quite satisfactory.
It proves that the dynamical spin accumulation builds up in the molecule
for frequencies $\omega \gtrsim \omega_{D>0}^\ast$.

The picture developed above changes significantly if a molecule is characterized by the uniaxial magnetic anisotropy of the easy-plane type~(\mbox{$D<0$}), see the right column of Fig.~\ref{fig2}. In particular, the major difference is that in such a situation at frequencies~\mbox{$\omega\lesssim D$} the conductance~$\GAP(\omega)$ [see Fig.~\ref{fig2}(e)] always reaches the limit of unitary transport, that is, \mbox{$\GAP(\omega)=(1-p^2)G_0$}, which is a signature of the Kondo effect. It is clear from Figs.~\ref{fig2}(f)-(g) that this effect is not related to the dynamical spin accumulation, as at low frequencies (\mbox{$\omega\ll D$}) transport is fully determined only by the conductance components diagonal in spin, \mbox{$\GAP(\omega)\approx\sum_\sigma\GAP_{\sigma\sigma}(\omega)$}. Furthermore, it can be noticed that the components off-diagonal in spin are now characterized by  smaller threshold frequencies~$\omega^\ast$, and that their magnitudes are larger, compare Fig.~\ref{fig2}(g) with Fig.~\ref{fig2}(c).
This, in turn, affects values of the current spin polarization~$\SPAP(\omega)$,
which in the present case exceed those for a spin-isotropic molecule, see Fig.~\ref{fig2}(h).
However, in other respects, the behavior of the relevant quantities under discussion,
shown in Figs.~\ref{fig2}(e)-(h), qualitatively resembles that observed for~\mbox{$D>0$}.

To understand the origin of the Kondo effect revival we again invoke the spectrum of a free-standing molecule.
Since the molecular spin is characterized  by the uniaxial magnetic anisotropy of the easy-plane type, it means that the ground state doublet of the $S+1/2$~spin multiplet is formed by the states with the~lowest $S_z^\text{tot}$~component, namely, \mbox{$\ket{S_z^\text{tot}=\pm1/2}$}. These states arise as superpositions of states
\mbox{$
	\big\{
	\ket{\pm1/2}_\text{OL}
	\otimes
	\ket{0}_\text{core}
	,
	\ket{\mp1/2}_\text{OL}
	\otimes
	\ket{\pm1}_\text{core}
	\big\}
$}, 
from which it is clear that the ground state doublet can now support the electron spin exchange processes in the~OL \mbox{---the}~mechanism underlying the Kondo effect.
The effective exchange interaction between the molecule and 
the leads is conditioned by the excitation energies between the ground state doublet
and the empty and fully-occupied orbital-level molecular states, which basically depends on all model parameters
in a nontrivial fashion. Consequently, it is a tedious task to provide a simple analytical formula 
for the energy scale~$\omega^*$. Instead, let us just
conclude from the inspection of frequency-dependent transport characteristics
shown in the right column of Fig.~\ref{fig2}
that the Kondo temperature is of the order of the magnetic anisotropy constant,
\mbox{$\TK\sim |D|$}, while the energy scale~\mbox{$\omega_{D<0}^\ast$}
is slightly smaller than~$\TK$ and grows linearly with $|D|$.
This behavior can be clearly seen in the dynamical spin accumulation
shown in Fig.~\ref{fig2}(g). The onset of~$\GAP_{\uparrow\downarrow}(\omega)$
moves to larger frequencies with increasing~$|D|$ 
and in the limit of very large magnetic anisotropy the system's dynamical
behavior approaches the quantum dot case, indicated by the thin dotted line.

From the above discussion one can already formulate some more universal 
statements concerning the behavior of the dynamical spin accumulation.
It is clear that this effect is most effective when the spin-exchange processes are relevant.
This happens for frequencies corresponding to the energy scale responsible
for the formation of the Kondo state.
Thus, one can observe that the maximum in 
$\GAP_{\sigma\overline{\sigma}}(\omega)$ develops for some resonant
frequency~$\omegar$, which is of the order of the Kondo temperature, \mbox{$\omegar \approx \TK$}.
On the other hand, the width of this maximum 
depends strongly on the frequency range of the slope 
when the conductance as a function of~$\omega$ increases due to the Kondo effect \mbox{---note} that we discuss the behavior on logarithmic scale.
This is why a broad maximum can be observed 
for spin-isotropic molecules, while for finite magnetic anisotropy
the frequency range of enhanced spin accumulation is much reduced.
When the conductance~$\GAP(\omega)$ reaches a plateau with lowering~$\omega$,
the spin-flip processes become quenched and a many-body
delocalized screened-spin state is formed between the molecule's spin and
the spins of conduction electrons.
The electrons, when tunneling through the junction,
experience then only a phase shift and spin-flip processes are suppressed~\cite{Hewson_book}.
As a consequence, 
the off-diagonal components of frequency-dependent conductance
get suppressed and the effect of dynamical spin accumulation disappears.
The energy scale when this happens is, in turn, described by~$\omega^*$,
which corresponds to the onset of dynamical spin accumulation
with increasing the driving frequency~$\omega$.

As far as the height of the maximum in $\GAP_{\sigma\overline{\sigma}}(\omega)$
is concerned, one can see that if the value of zero-frequency conductance 
is smaller than its maximum value, which effectively means that the Kondo effect
cannot fully develop in the system, the magnitude of $\GAP_{\sigma\overline{\sigma}}(\omega)$
gets reduced. This can be especially seen for the easy-axis type of magnetic anisotropy presented in  the left column of Fig.~\ref{fig2}.
On the other hand, for magnetic molecules with anisotropy of the easy-plane type,
the ground state is always a spin doublet,
so that at low frequencies
the Kondo effect can fully develop and, consequently, while the position 
of maximum in $\GAP_{\sigma\overline{\sigma}}(\omega)$ depends 
on $D$, its maximum value does not. In fact, the 
maximum value of the dynamical spin accumulation is then
comparable to the quantum dot case, see the right column of Fig.~\ref{fig2}.

\subsubsection{\label{sec:Uniaxial_and_transverse}Uniaxial and transverse magnetic anisotropy}
%
\begin{figure}[t]
	\includegraphics[scale=1.05]{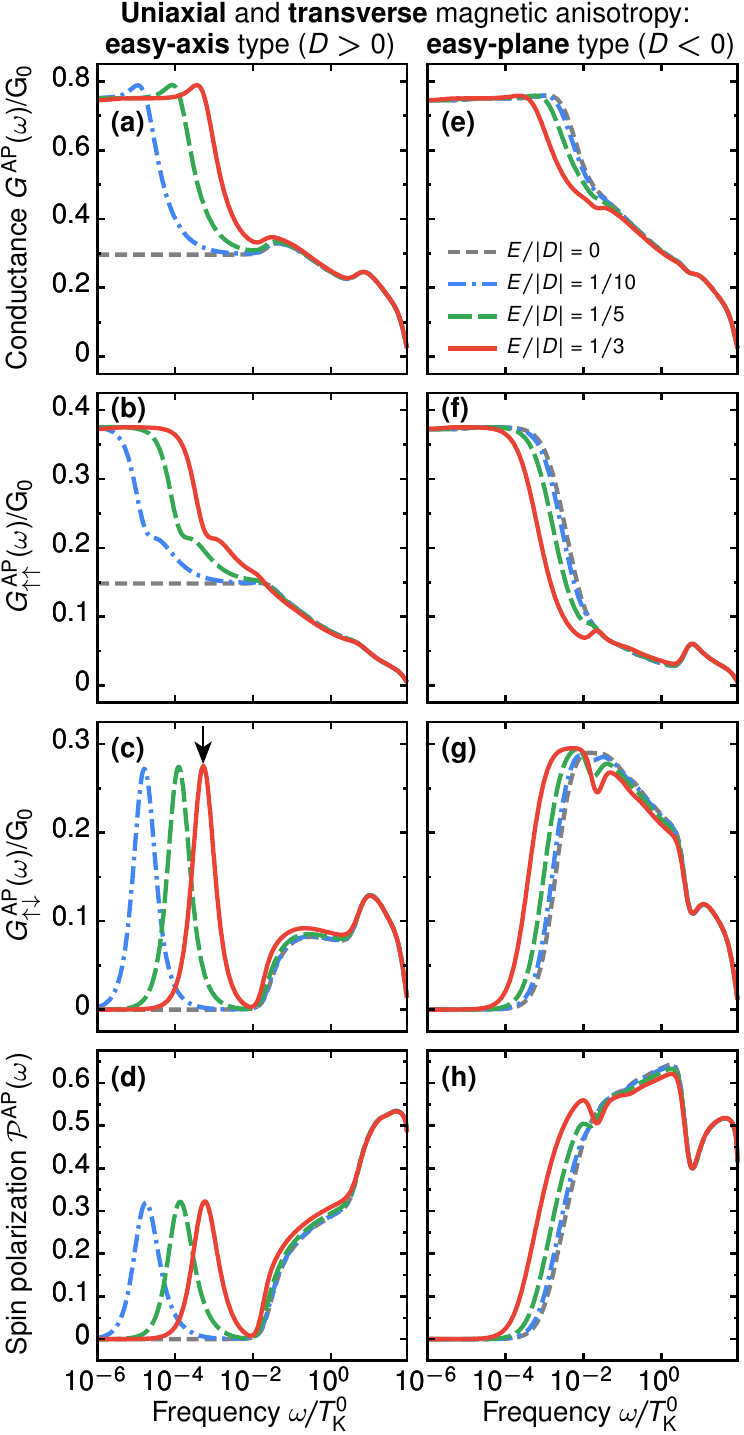}
	\caption{
	Analogous to Fig.~\ref{fig2} except that now also the effect of the \emph{transverse} component of magnetic anisotropy~$E$ is included for a selected value of uniaxial magnetic anisotropy constant \mbox{$|D|/\TKqd=10^{-2}$}.
	Here, the transverse constant~$E$ is always assumed positive.
	Note that to enable an easy comparison with results in Fig.~\ref{fig2}, the scales are kept identical as in Fig.~\ref{fig2}, and the finely dashed line representing the case of~$E=0$ is added to serve as the reference line.
    \label{fig3}
    }
\end{figure}

The uniaxial component of magnetic anisotropy along the $z$~axis is often accompanied by the transverse one, described by the term~$E\big(\opS_x^2-\opS_y^2\big)$ in Eq.~(\ref{eq:H_core}). In essence, it captures the effect of breaking the rotational symmetry around the $z$~axis, which, in other words, means that the internal spin has a tendency to align along some directions with respect to the plane perpendicular to the~$z$~axis. In particular, for~$E>0$ the energy of the spin is minimized if it is oriented along the~$y$~axis.
The significance of this transverse term of magnetic anisotropy lies in the fact that such a term leads to mixing of the axial spin states~\mbox{$\ket{S_z}_\text{core}$}, which can be easily seen if one introduces in Eq.~(\ref{eq:H_core}) the ladder operators
\mbox{$
	\opS_\pm
	\equiv
	\opS_x\pm i\opS_y
	.
$
}
Generally, this mixing is at the foundation of many important effects influencing transport, such as, the quantum tunneling of spin~\cite{Romeike_Phys.Rev.Lett.96/2006,Misiorny_Phys.Rev.B90/2014}, or the Berry-phase blockade~\cite{Leuenberger_Phys.Rev.Lett.97/2006,Gonzalez_Phys.Rev.Lett.98/2007,Gonzalez_Phys.Rev.B78/2008}.

Figure~\ref{fig3} illustrates how inclusion of the transverse magnetic anisotropy~(\mbox{$D\neq0$} and \mbox{$E\neq0$}) affects the finite-frequency conductance and the dynamical spin accumulation in the case of the easy-axis~(\mbox{$D>0$}, left column) and easy-plane~(\mbox{$D<0$}, right column) type of uniaxial magnetic anisotropy. 
Let us first focus on the case of~\mbox{$D>0$}.
The first noticeable difference, as compared with the case of~\mbox{$D>0$} and~\mbox{$E=0$} [see Figs.~\ref{fig2}(a)-(d)], is that one observes the revival of the Kondo effect for sufficiently low frequencies. Such a restoration of transport occurs as a consequence of the mixing caused by the second term of Hamiltonian~(\ref{eq:H_core}), because now each molecular spin state effectively  becomes a superposition of all possible OL electronic spin and internal spin states~\cite{Misiorny_Phys.Rev.B90/2014}. This, in turn, means that the spin exchange processes leading to transitions between the states of the ground state doublet are permitted.
Furthermore, for $\omega \approx \omegar$ a pronounced maximum in the dynamical conductance~\mbox{$\GAP(\omega)$} is visible, Fig.~\ref{fig3}(a), and it stems from the dynamical spin accumulation~$\GAP_{\uparrow\downarrow}(\omega)$, which at this particular frequency~$\omegar$ exhibits a sharp resonance, as one can see in Fig.~\ref{fig3}(c). We also note that at~$\omegar$ a kink develops in the conductance component diagonal in spin~$\GAP_{\uparrow\uparrow}(\omega)$, Fig.~\ref{fig3}(b),
while spin polarization $\mathcal{P}^{AP}(\omega)$ exhibits a local maximum, 
Fig.~\ref{fig3}(d).

\begin{figure}[t]
	\includegraphics[width=1\columnwidth]{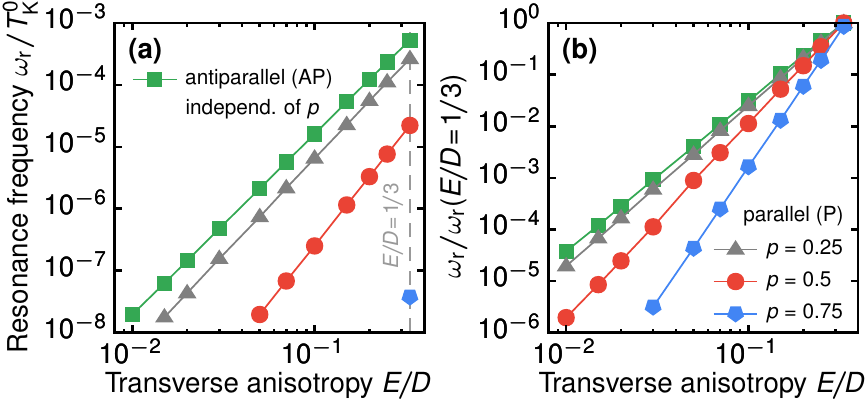}
	\caption{
	(a) Dependence of the position~$\omegar$ of the resonance
	in the conductance component off-diagonal in spin,~$G_{\sigma\overline{\sigma}}(\omega)$, on the transverse magnetic anisotropy constant~$E$. 
	For the antiparallel (AP) magnetic configuration (squares) the data point for~\mbox{$E/D=1/3$} (marked by a finely dashed vertical line) corresponds to the resonance indicated in Fig.~\ref{fig3}(c) by the arrow. Note that for this magnetic configuration~$\omegar$ is independent of the spin polarization~$p$, see Fig.~\ref{fig5}(g), and that lines connecting the data points serve as a guide for eyes.
	The data points at~\mbox{$E/D=1/3$} for the parallel (P) magnetic configuration represent the position of relevant peaks in Fig.~\ref{fig7}(g). 
	(b) Data points shown in (a) rescaled by~\mbox{$\omegar(E/D=1/3)$} to highlight the change of the slope occurring  in the parallel magnetic configuration.
    Here, \mbox{$D/\TKqd=10^{-2}$} and other parameters as in Fig.~\ref{fig3}.
    \label{fig4}
    }
\end{figure}

The behavior of dynamical transport properties in the presence of transverse component of magnetic anisotropy can be understood by invoking the discussion in the previous section. Generally, one can again notice that the value of~$\omegar$ coincides with the Kondo temperature, that is,  the dynamical spin accumulation exhibits a maximum at~\mbox{$\omega = \omegar \approx \TK$}. Such a feature arises for all nonzero values of anisotropy~$E$ considered in the figure, and importantly, the position of this feature  depends strongly on the transverse anisotropy component \mbox{---small} changes in~$E$ lead to a large shift of the resonance frequency~$\omegar$.
This is directly related to a strong dependence of the Kondo temperature on the model parameters and, in particular, finite transverse anisotropy \cite{Misiorny_Phys.Rev.B90/2014}. An explicit, numerically determined dependence of~$\TK$ on~$E$ can be seen in Fig.~\ref{fig4}, which illustrates how the position of the resonance in the dynamical spin accumulation~$\GAP_{\sigma\overline{\sigma}}(\omega)$ evolves when the transverse anisotropy parameter~$E$ is modified. 
Clearly, small changes in~$E$ result in large  modification of~$\omegar$ and, thus,~$\TK$. From the slope of the calculated curve, see squares in Fig.~\ref{fig4}, we estimate that~\mbox{$\omegar \propto (E/D)^{-3}$}. We also note that the position of this curve is insensitive to the spin polarization~$p$ of electrodes.
Moreover, it can be observed in Fig.~\ref{fig3}(c) that the width of the peak in~$\GAP_{\uparrow\downarrow}(\omega)$ and $\SPAP(\omega)$ \mbox{---plotted} on a logarithmic \mbox{scale---} hardly depends on~$E$. One can  conclude, thus, that the width of the resonance in dynamical spin accumulation, which occurs at~\mbox{$\omegar\approx \TK$},
is also approximately given by the Kondo temperature.

On the contrary, in the case of the uniaxial magnetic anisotropy of the easy-plane type~(\mbox{$D<0$}),
presented in the right column of Fig.~\ref{fig3}, the dynamical conductance~$\GAP(\omega)$ is modified more subtly. Now, one observes the well-developed Kondo effect [Figs.~\ref{fig3}(e)-(f)] and a pronounced maximum both in the dynamical spin accumulation~$\GAP_{\uparrow\downarrow}(\omega)$ [Fig.~\ref{fig3}(g)] and in the current spin polarization~$\SPAP(\omega)$ [Fig.~\ref{fig3}(h)], already visible in the absence of transverse component of magnetic anisotropy. 
The increase of~$E$ results only in a small reduction of the threshold frequency~$\omega^\ast$ at which $\GAP_{\uparrow\downarrow}(\omega)$ starts building up, and at which also the suppression of the Kondo effect takes place \mbox{---in} other words, the raise of~$E$ leads to a slight decrease of the Kondo temperature. Moreover, for larger~$E$ the maximum in~$\GAP_{\uparrow\downarrow}(\omega)$ gets broader and eventually a small dip on the top of it develops. Interestingly, at the frequency where this dip is observed, one can also notice a local maximum in~$\GAP_{\uparrow\uparrow}(\omega)$, see Fig.~\ref{fig3}(f).

\subsection{\label{sec:Spin_polarization}Influence of the spin polarization of electrodes}
%

\begin{figure}[t]
	\includegraphics[scale=1.05]{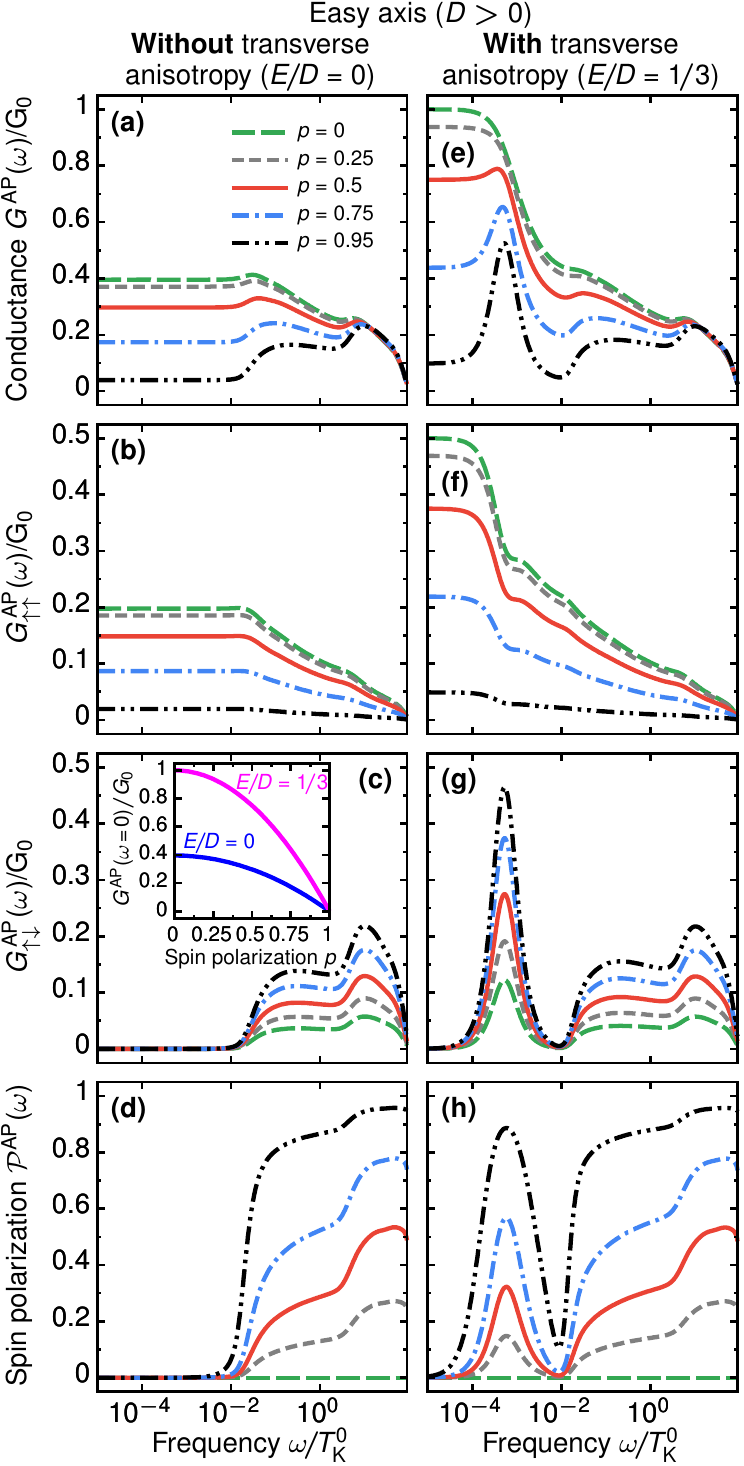}
	\caption{
	Evolution of the frequency-dependent conductance~$\GAP(\omega)$ in~(a,e), its spin components~$\GAP_{\uparrow\uparrow}(\omega)$ in~(b,f) and~$\GAP_{\uparrow \downarrow}(\omega)$ in~(c,g), as well as the current spin polarization~$\SPAP(\omega)$ in~(d,h) presented as a function of the spin-polarization coefficient~$p$ of electrodes for~\mbox{$D/\TKqd=10^{-2}$} (the uniaxial magnetic anisotropy of the easy-axis type).
	\emph{Left} (\emph{right}) \emph{column} corresponds to the case without (with) the transverse component of the magnetic anisotropy included, that is, for~\mbox{$E=0$} (\mbox{$E=D/3$}).
	The inset in (c) presents the dependence of the zero-frequency conductance~\mbox{$\GAP(\omega=0)$} on the spin polarization~$p$ in the case of~\mbox{$E=0$} and~\mbox{$E/D=1/3$}.
    Other parameters are the same as in Fig.~\ref{fig2}.
    \label{fig5}
    \vspace*{-10pt}
    }
\end{figure}

\begin{figure}[t]
	\includegraphics[scale=1.05]{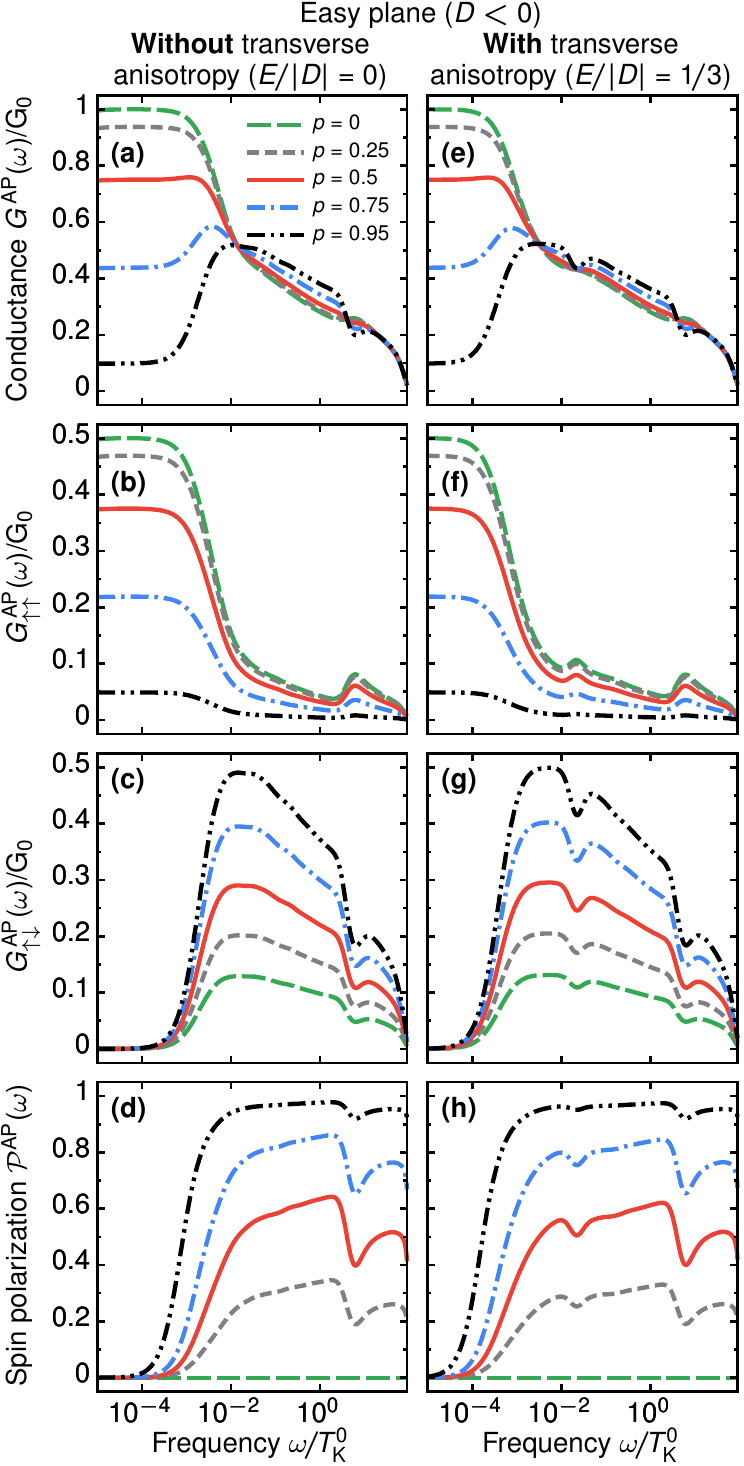}
	\caption{
	Analogous to Fig.~\ref{fig5} except that here the case of the uniaxial magnetic anisotropy of the easy-plane type~($D/\TKqd=-10^{-2}$) is considered.
	Note that the dependence of the zero-frequency dynamical conductance~\mbox{$\GAP(\omega=0)$} on~$p$ is the same regardless of whether the transverse component of magnetic anisotropy is present or not, and it is given by the curve for~$E/D=1/3$ shown in the inset to Fig.~\ref{fig5}(c).
    \label{fig6}
    }
\end{figure}

In order to explore further the subtle interplay between the Kondo effect
and the dynamical spin accumulation, here we consider the effect of spin polarization of electrodes. 
For this purpose, in Fig.~\ref{fig5} and Fig.~\ref{fig6} we show how the dynamical transport response of the system under investigation changes for different values of the spin polarization parameter~$p$ in the case of \mbox{$D>0$}~(Fig.~\ref{fig5}) and \mbox{$D<0$}~(Fig.~\ref{fig6}), respectively. The left (right) column in those figures corresponds to the case of zero (finite) transverse magnetic anisotropy constant $E$.

Let us first consider the case of~\mbox{$D>0$} shown in Fig.~\ref{fig5}. Generally, as expected for the Kondo effect, with the increase of~$p$ the low-frequency conductance~\mbox{$\GAP(\omega\lesssim\omega^\ast)$}, that is, below the threshold frequency~$\omega^\ast$ for the dynamical spin accumulation to kick in, becomes suppressed, see Figs.~\ref{fig5}(a,e).
This behavior stems from the fact that in electrodes characterized by a large degree of spin polarization, there is a great imbalance between the numbers of spin-majority and spin-minority electrons to be involved in the spin exchange processes leading to the Kondo effect. In consequence, the larger the imbalance is, the less effective these processes become, and the more the Kondo effect becomes suppressed.
The dependence of the zero-frequency conductance~\mbox{$\GAP(\omega=0)$} in the antiparallel magnetic configuration on the spin polarization $p$ of electrodes is presented in the inset to Fig.~\ref{fig5}(c) and it can be described by a simple formula, 
$
	\mbox{$\GAP(\omega=0)$}
	\equiv
	\mbox{$\GAP(\omega=0,p)$}
	= 
	\mbox{$(1-p^2)\GAP(\omega=0,p=0)$}
$.

The situation changes qualitatively for frequencies~\mbox{$\omega\gtrsim\omega^\ast$}, where the off-diagonal-in-spin component of conductance~$\GAP_{\uparrow \downarrow}(\omega)$, Figs.~\ref{fig5}(c,g), starts contributing significantly, so that the dynamical spin accumulation emerges as the dominant effect.
Importantly, although one can notice that~$\GAP_{\uparrow \downarrow}(\omega)$ does not vanish even when the electrodes are non-magnetic (\mbox{$p=0$}), no spin polarization of the current is observed in such a case, that is, $\SPAP(\omega)=0$, as one can see in Figs.~\ref{fig5}(d,h). Moreover, the behavior of the dynamical conductance~$\GAP(\omega)$ is then primarily governed by its diagonal-in-spin component~$\GAP_{\uparrow \uparrow}(\omega)$ \mbox{---compare} in Figs.~\ref{fig5} panels~(a,e) with~(b,f).
On the other hand, in the opposite limit of strongly spin-polarized electrodes~(\mbox{$p>0.5$}), where~$\GAP_{\uparrow \uparrow}(\omega)$ gets progressively suppressed for large~$p$, the features associated with the dynamical spin accumulation~$\GAP_{\uparrow \downarrow}(\omega)$ become in fact increasingly visible in the total conductance~$\GAP(\omega)$ within the entire range of frequencies~$\omega$ \mbox{---compare} in Figs.~\ref{fig5} panels~(a,e) with~(c,g).
This observation illustrates the key generic difference between the response of the Kondo effect and the dynamical spin accumulation to a large spin polarization of electrodes. Specifically, unlike the Kondo effect, the dynamical spin accumulation is augmented with the increase of~$p$. This is a direct consequence of the fact that $\GAP_{\uparrow \downarrow}(\omega)$ is associated with majority spin bands of both leads, while the diagonal-in-spin components~$\GAP_{\sigma\sigma}(\omega)$ depend on both majority and minority spin bands.

In addition, one can note that while the low (\mbox{$\omega \ll \omega^*$}) and high-frequency (\mbox{$\omega \gg \omega^*$}) behavior of the dynamical conductance and its spin-resolved contributions is qualitatively similar in the case of~\mbox{$E=0$} and finite~$E$, huge differences occur when~\mbox{$\omega \approx \omega^*$}. 
For~\mbox{$E=0$}, the dynamical spin accumulation starts growing for \mbox{$\omega \gtrsim \omega^*$} to reach a plateau, whereas for finite $E$, $\GAP_{\uparrow \downarrow}(\omega)$ increases to form a strong maximum, the height of which grows with increasing~$p$. 
To understand this difference let us recall that  in the absence of transverse magnetic anisotropy the Kondo effect develops only partially. This implies that dynamical spin accumulation has a moderate, relatively wide in frequencies, maximum. On the other hand, for finite transverse magnetic anisotropy the Kondo resonance can fully develop with a clear sharp maximum in dynamical spin accumulation at~\mbox{$\omegar \approx \TK$}, which becomes greatly enhanced with increasing spin polarization $p$. In fact, in the limit of half-metallic leads (\mbox{$p=1$}), the dynamical conductance would be exclusively due to the effect of dynamical spin accumulation, that is,
\mbox{$
	\GAP(\omega)
	=
	\sum_\sigma
	\GAP_{\sigma\overline{\sigma}}(\omega)
$}.
We also note that while the spin-resolved components of the frequency-dependent conductance strongly depend on~$p$, the characteristic energy scales, $\omega^*$, $\TK$ and consequently $\omegar$, hardly do so, see Fig.~\ref{fig5}.

The above discussion is also relevant to the case of  easy-plane type of magnetic anisotropy (\mbox{$D<0$}), which is shown in Fig. \ref{fig6}. With raising the spin polarization, the diagonal-in-spin components of the dynamical conductance become suppressed [Figs. \ref{fig6}(b,f)], while the off-diagonal component~$\GAP_{\uparrow \downarrow}(\omega)$ increases [Figs. \ref{fig6}(c,g)], and for sufficiently large $p$ gives a dominant contribution to the total conductance.
In such a situation, the interplay of these two contributions results in a non-monotonic frequency dependence of~$\GAP(\omega)$, see Figs.~\ref{fig6}(a,e), which exhibits a local maximum due to dynamical spin accumulation. Similar to the case of easy-axis magnetic anisotropy presented in Fig.~\ref{fig5}, large spin polarization $p$ of the leads induces large spin polarization of the current, see Figs.~\ref{fig6}(d,h). As far as the effects related to the transverse component of magnetic anisotropy are concerned, finite $E$ results mainly in quantitative changes in the dynamical response of the system, cf. the left and right column of Fig.~\ref{fig6}, manifesting as a small reduction of the Kondo temperature and, consequently, the energy scale~$\omega^*$. However, a qualitative difference can be still observed in the dynamical
spin accumulation, which in the case of \mbox{$E=|D|/3$}
exhibits a small dip at intermediate frequencies, see 
Fig.~\ref{fig6}(g) for $\omega/\TKqd \approx 10^{-2}$.

\subsection{\label{sec:Parallel}Parallel magnetic configuration of the junction}

\begin{figure}[t]
	\includegraphics[scale=1.05]{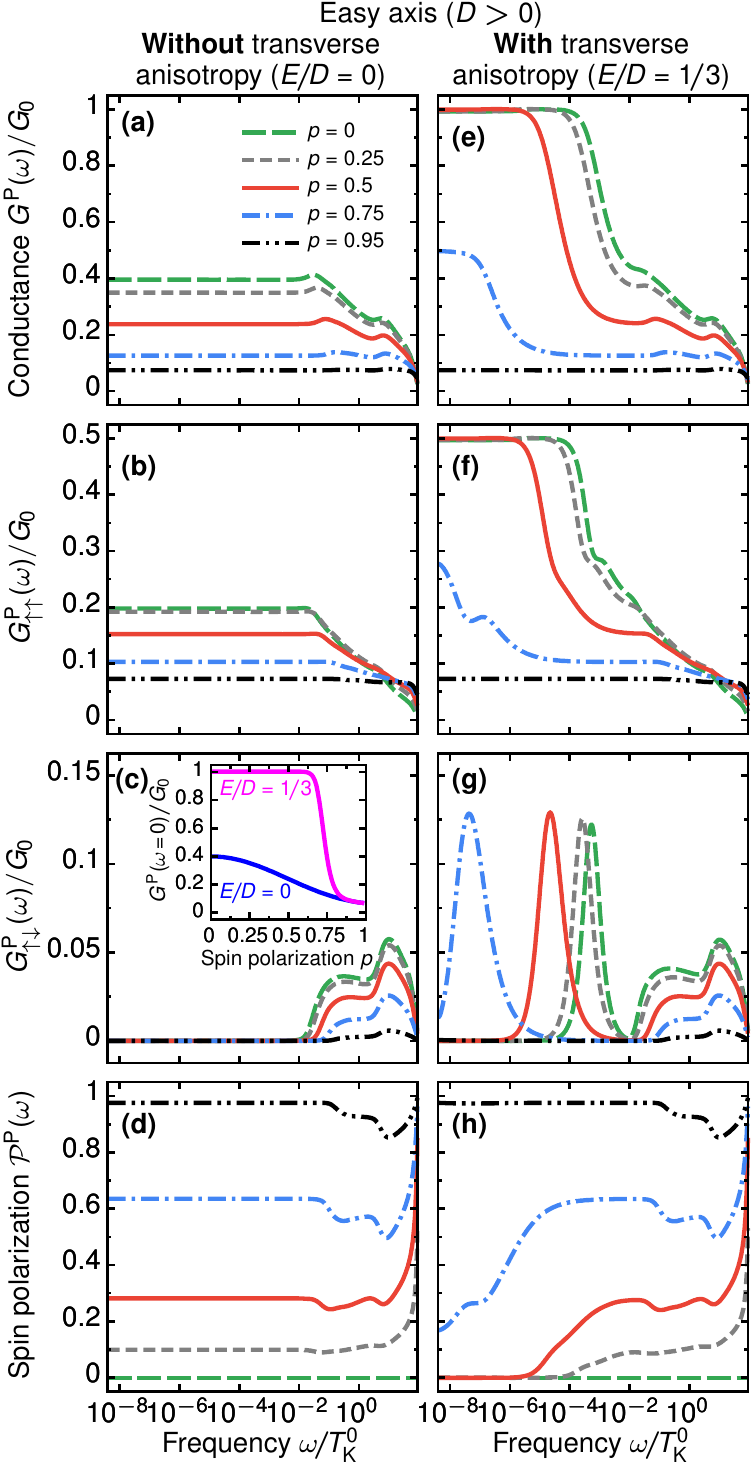}
	\caption{
	Analogous to Fig.~\ref{fig5} but now the \emph{parallel} (P) magnetic configuration of the junction is shown.
	Note that at present \mbox{$\GP_{\uparrow\downarrow}(\omega)=\GP_{\downarrow\uparrow}(\omega)$}, and thus, \mbox{$\SP_\text{dsa}(\omega)=0$}, so that the current spin polarization occurs only due to the diagonal-in-spin terms~\mbox{$\GP_{\uparrow\uparrow}(\omega)\neq\GP_{\downarrow\downarrow}(\omega)$}, that is, \mbox{$\SPP(\omega)\equiv\SPP_0(\omega)$}.
	Note that although, for the sake of consistency, we plot here the same set of $p$~values as in previous figures, the frequency range has been extended here to include lower values of~$\omega$.   
	Moreover, the long-dashed line is identical to that in Fig.~\ref{fig5} and it serves as the reference line.
    Recall that \mbox{$D/\TKqd=10^{-2}$} and the other parameters are the same as in Fig.~\ref{fig2}.
    \label{fig7}
    }
\end{figure}
\begin{figure}[t]
	\includegraphics[scale=1.05]{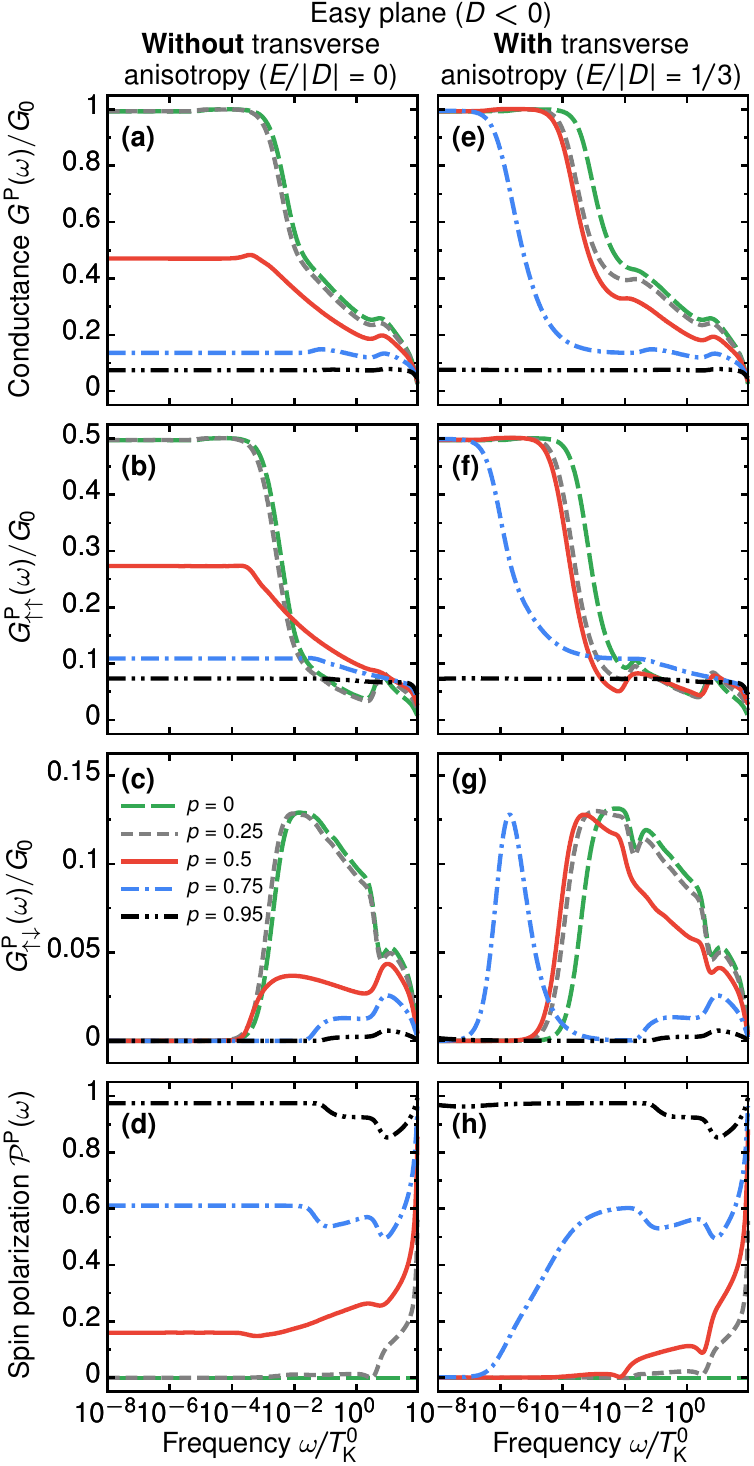}
	\caption{
	Analogous to Fig.~\ref{fig7} except that here the case of the uniaxial magnetic anisotropy of the easy-plane type~($D/\TKqd=-10^{-2}$) is considered.
    The other parameters are the same as in Fig.~\ref{fig2}.
    \label{fig8}
    }
\end{figure}

Up to this point, the discussion has been concentrated on the situation of the junction in the antiparallel magnetic configuration, which allowed us to exclude from the picture some subtle effects due to the spintronic effective exchange fields.
To acquire a complete understanding of how the magnetic configuration of the junction affects the process of dynamical spin accumulation, we now also consider the parallel magnetic configuration of the junction.

For this purpose, we present in Figs.~\ref{fig7} and \ref{fig8}  how the dynamical transport characteristics of the system depend on the spin polarization~$p$ of electrodes for the  parallel magnetic configuration. 
To begin with, let us first focus on the case of the easy-axis type of uniaxial magnetic anisotropy (\mbox{$D>0$}) shown in Fig.~\ref{fig7}.
One can see that if only the uniaxial component of magnetic anisotropy is present the dynamical conductance~$\GP(\omega)$ does not differ qualitatively from the antiparallel case, compare Fig.~\ref{fig7}(a) with Fig.~\ref{fig5}(a). Nevertheless, two key quantitative differences can be spotted immediately: 
First, the different values of conductance in the zero-frequency limit, \mbox{$\GAP(\omega=0)\geqslant\GP(\omega=0)$} for \mbox{$0<p\lesssim0.85$} [compare the insets in Fig.~\ref{fig5}(c) and Fig.~\ref{fig7}(c)], 
and second, the behavior of the resonance around the threshold frequency~$\omega^\ast$ which marks the onset of the dynamical spin accumulation, as discussed in~Sec.~\ref{sec:Uniaxial}. Specifically, with the increase of the spin polarization~$p$ of electrodes this resonance becomes shifted towards larger frequencies and its magnitude gets attenuated.

The origin of the suppression of~\mbox{$\GP(\omega=0)$} and the shift can be explained by taking into account the fact that in the present configuration the spintronic effective exchange fields can arise. Since we consider that the molecule is tuned to the particle-hole symmetry point~(\mbox{$\varepsilon/U=-0.5$}), one expects actually only the quadrupolar field~\cite{Misiorny_NaturePhys.9/2013}, which essentially provides an additional uniaxial contribution~$-D_\text{s}^{}\big(\opS_z+\ops_z\big)^2$ (with~\mbox{$D_\text{s}^{}>0$}) to the magnetic anisotropy Hamiltonian~(\ref{eq:H_core}). As a result, the~energy separation~\mbox{$\Delta\propto D+D_\text{s}$} between the states participating in transport, that is, the ground state doublet and first excited doublet in the \mbox{$S+1/2$} spin multiplet, increases. This, in turn, translates into the larger threshold frequency~$\omega_{D>0}^\ast$ and also means that the spin exchange processes leading to the Kondo effect at low frequencies are more subdued, as compared with the antiparallel case where~\mbox{$D_\text{s}=0$}.

On the other hand, the suppression of the features occurring for~\mbox{$\omega\gtrsim\omega^\ast$} has its roots in the response of the dynamical spin accumulation~$\GP_{\uparrow\downarrow}(\omega)$ to increasing~$p$,
see Fig.~\ref{fig7}(c). Importantly, this response is strikingly different
from that for the antiparallel magnetic configuration in Fig.~\ref{fig5}(c).
First of all, it should be noted that now one finds~\mbox{$\GP_{\uparrow\downarrow}(\omega)=\GP_{\downarrow\uparrow}(\omega)$}, which straightforwardly leads to the conclusion that the dynamical spin accumulation does not contribute to the spin polarization of the current, \mbox{$\SPP_\text{dsa}(\omega)=0$}. In fact, the current spin polarization~$\SPP(\omega)$ shown in Fig.~\ref{fig7}(d) is exclusively due to the difference between the diagonal-in-spin components~$\GP_{\uparrow\uparrow}(\omega)$ and~$\GP_{\downarrow\downarrow}(\omega)$, that is, \mbox{$\SPP(\omega)\equiv\SPP_0(\omega)$}.
Moreover, the intensity of~$\GP_{\uparrow\downarrow}(\omega)$ is significantly
reduced with respect to~$\GAP_{\uparrow\downarrow}(\omega)$,
and it exhibits the opposite behavior with the increase of~$p$, namely,
in the parallel magnetic configuration the dynamical spin accumulation is diminished for large~$p$.
This behavior can be understood by realizing that 
in the antiparallel configuration the off-diagonal conductance
$\GAP_{\uparrow\downarrow}(\omega)$ is associated with the majority-spin
subbands of both leads. Consequently, increasing the spin polarization
results in an enhancement of dynamical spin accumulation.
On the other hand, in the parallel configuration, the off-diagonal
components depend on both majority and minority spin subbands of both leads,
such that the minority spin channel provides a bottleneck for 
$\GP_{\uparrow\downarrow}(\omega)$.
Consequently, in the parallel magnetic configuration
the effect of dynamical spin accumulation becomes suppressed with increasing $p$,
contrary to the case of antiparallel configuration.
Note also that in the limit of half-metallic leads ($p\to 1$),
$\GP_{\uparrow\downarrow}(\omega)$ would be fully suppressed
and the total conductance would be exclusively given by the majority spin
diagonal component of the conductance, while 
in the antiparallel configuration all components would disappear,
except for $\GAP_{\uparrow\downarrow}(\omega)$,
namely,  the total conductance would be
only due to the dynamical spin accumulation.

\begin{figure*}[t]
	\includegraphics[width=\textwidth]{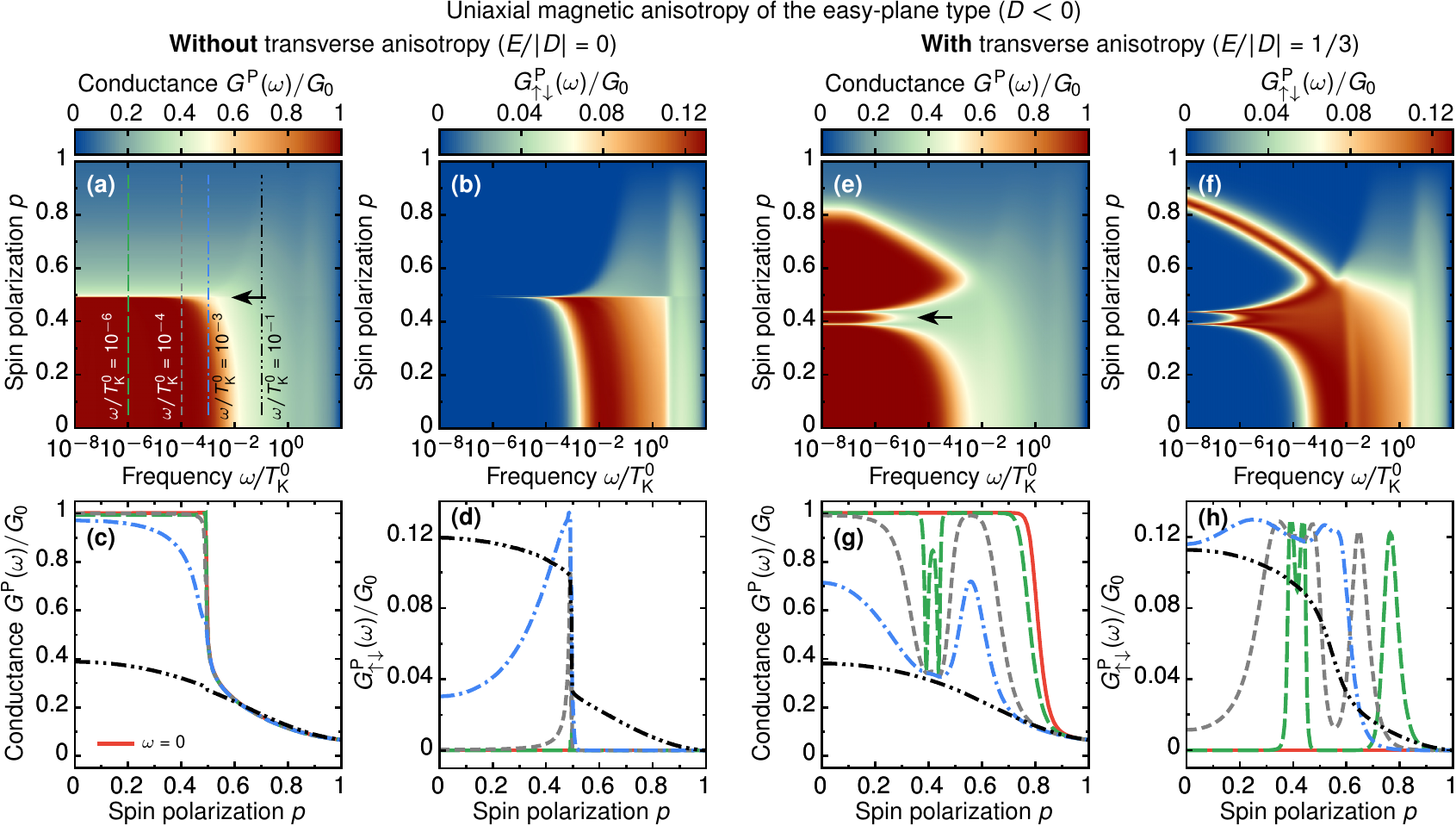}
	\caption{
	Evolution of the dynamical conductance~$\GP(\omega)$ and its off-diagonal-in-spin component~$\GP_{\uparrow\downarrow}(\omega)$ as functions of the spin polarization~$p$ of electrodes for the uniaxial magnetic anisotropy of the easy-plane type~(\mbox{$D/\TKqd=-10^{-2}$}). 
	\emph{Left panels}~[(a)-(d)] correspond to the case without the transverse component of magnetic anisotropy~(\mbox{$E=0$}), whereas \emph{right panels}~[(e)-(f)] to the case when the transverse component is included~(\mbox{$E/|D|=1/3$}).
	\emph{Bottom row} presents cross-sections of the respective map plots in the top row for chosen values of frequencies~$\omega$ marked in~(a), with the solid line standing for the zero-frequency limit.
	The other parameters are the same as in Fig.~\ref{fig2}.
	\label{fig9}
	}
\end{figure*}
\begin{figure}[t]
	\includegraphics[scale=1.05]{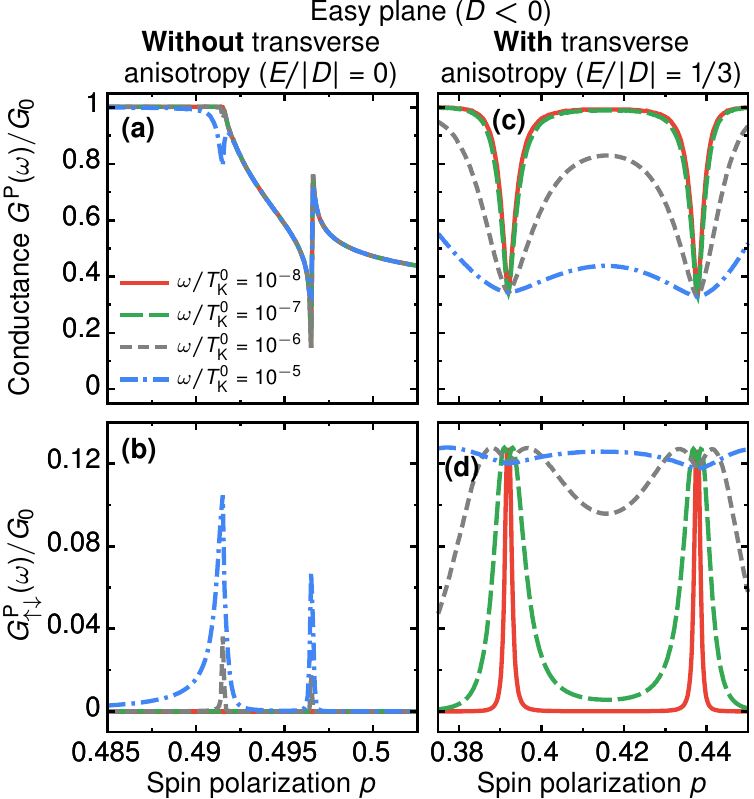}
	\caption{
	(a)-(b) [(c)-(d)] Cross-sections of the map plots shown in the top row of Fig.~\ref{fig9} for selected values of frequencies~$\omega$ and resolved around the features marked by the arrow in Fig.~\ref{fig9}(a)~[(e)].
	\label{fig10}
	}
\end{figure}

Let us now take the transverse component of magnetic anisotropy
into consideration, see the right column of Fig.~\ref{fig7}. 
Comparing with the case of the antiparallel magnetic configuration shown in Fig.~\ref{fig5}(e),
one observes in Fig.~\ref{fig7}(e) that the suppression of the Kondo effect  occurring  for large~$p$
proceeds now in a qualitatively different manner. With the increase of~the spin polarization,
the Kondo temperature~$\TK$ initially decreases whereas
the zero-frequency conductance~\mbox{$\GP(\omega=0)=G_0$} remains unaffected,
and only above some threshold value of~$p$ also~\mbox{$\GP(\omega=0)$}
becomes diminished \mbox{---for} a detailed evolution of~\mbox{$\GP(\omega=0)$} as a function of~$p$ see the inset in Fig.~\ref{fig7}(c).
In consequence, as long as the Kondo effect dominates transport,
the current injected into the right electrode does not display
the spin polarization, that is, \mbox{$\SPP(\omega)=0$} for $\omega \lesssim \TK$, see Fig.~\ref{fig7}(h).
Furthermore, we notice that  in the large frequency regime~\mbox{$\omega\gtrsim\omega^\ast$}, the transport features due to the dynamical spin accumulation behave identically to those discussed in the situation without the transverse component of magnetic anisotropy. 
On the other hand, in the opposite limit it can be seen that, unlike for the antiparallel configuration, the resonance arising in~$\GP_{\uparrow\downarrow}(\omega)$ shifts its position~$\omega_r$ towards smaller frequencies when~$p$ becomes larger. This results from the dependence of the Kondo temperature on~$p$, which decreases as $p$ raises.

The explicit dependence of $\omega_r$ on the transverse  magnetic anisotropy is presented in Fig.~\ref{fig4} for a couple of selected values of spin polarization. It can be clearly seen that the resonance frequency, and, thus, also the Kondo temperature, strongly depends now both  on the value of $E$ and the spin polarization~$p$. Interestingly, the slope of  the dependence~of~$\omega_r$ on~$E/D$ differs slightly from the case of antiparallel configuration, and it is also sensitive to a change of~$p$, see Fig.~\ref{fig4}(b). Note that the data points (squares) in Fig.~\ref{fig4} corresponding to the antiparallel magnetic configuration are actually valid also for the nonmagnetic junction (\mbox{$p=0$}).
One can see that with increasing $p$, the dependence of the Kondo temperature
on $E$ becomes sharper. Moreover, because the Kondo temperature 
greatly depends on spin polarization, for $p\gtrsim 0.75$,
the Kondo effect actually develops at extremely small energy scales,
which are completely not relevant from the experimental point of view.

The behavior of dynamical system's response changes
if one considers the uniaxial magnetic anisotropy of the easy-plane type~(\mbox{$D<0$}), see~Fig.~\ref{fig8}.
It can be seen that in such a case, already without the transverse component of magnetic anisotropy,
a significant qualitative difference in evolution of the dynamical conductance
as a function of~$p$ arises between the antiparallel~(left column in~Fig.~\ref{fig6})
and parallel~(left column in~Fig.~\ref{fig8}) magnetic configuration of the junction. 
One can notice that, unlike for the antiparallel case,
the transition between the highly conducting state for small~$p$
and the weakly conducting state for large~$p$ occurring for \mbox{$\omega\lesssim\omega^\ast$}
is rather abrupt.
Furthermore, also when including the transverse component of
magnetic anisotropy (right column in Fig.~\ref{fig8}),
this transition proceeds in a qualitatively different manner as compared
to the antiparallel case (right column in Fig.~\ref{fig6}).
Importantly, it can be observed that above some threshold value of~$p$
the dynamical spin accumulation~$\GP_{\uparrow\downarrow}(\omega)$
shown in Fig.~\ref{fig8}(g) changes its character and it develops
a resonance typical to the easy-axis type of uniaxial magnetic anisotropy, see Fig.~\ref{fig7}(g).

It is worth noting that the difference in the behavior of the dynamical conductance
observed in the parallel and antiparallel magnetic configurations on the value of the spin polarization of electrodes
has its origin in completely different mechanisms governing
the suppression of the Kondo effect. In the case of 
antiparallel configuration, the system 
behaves effectively as if coupled to nonmagnetic leads
and the effect of spin polarization enters only through the prefactors
in appropriate formulas. In the parallel configuration, on the other hand,
the effective couplings do depend on spin,
which results in nontrivial spin-resolved molecule-bath renormalization 
effects that give rise to finite local exchange fields.
As a consequence, the interplay between the degree of spin polarization,
which conditions the strength of exchange fields,
and the electronic correlations driving the Kondo effect,
gives rise to large qualitative differences, as compared to the case of antiparallel configuration.

In order to gain a better insight into how the dynamical conductance evolves with increasing the spin polarization~$p$, in Fig.~\ref{fig9} we analyze in detail the dependence of the total dynamical conductance~$\GP(\omega)$ and its off-diagonal-in-spin component~$\GP_{\uparrow\downarrow}(\omega)$ on~$p$. 
Indeed, we find for~\mbox{$E=0$}  [Figs.~\ref{fig9}(a)-(d)] that in the parallel magnetic configuration the transport response of the system is very sensitive to the change of~$p$, with  the suppression of the Kondo effect taking place suddenly at~\mbox{$p_0\approx0.5$} [Fig.~\ref{fig9}(a)] and  the value of~$\GP(\omega)$  altering for~$p>p_0$ only slightly, see the relevant cross-sections given by the dashed lines in Fig.~\ref{fig9}(c). 
As discussed above, the dynamical spin accumulation [Fig.~\ref{fig9}(b)] arises only for~\mbox{$\omega\gtrsim\omega^\ast$}, but at present the threshold energy~$\omega^\ast$ depends on~$p$ non-monotonically. Namely, it first decreases as~$p$ grows, while above~$p_0$ the opposite trend is visible, and~$\GP_{\uparrow\downarrow}(\omega)$ becomes quickly attenuated. 
In the left column of Fig.~\ref{fig10} we present how the total dynamical conductance~$\GP(\omega)$ [Fig.~\ref{fig10}(a)] and its off-diagonal-in-spin component~$\GP_{\uparrow\downarrow}(\omega)$ [Fig.~\ref{fig10}(b)] vary in~$p$ around~$p_0$. One can see that, in fact, $\GP(\omega)$ does not change smoothly during the transition between \mbox{$p\lesssim p_0$} and \mbox{$p\gtrsim p_0$} and two characteristic features arise \mbox{---also} two pronounced peaks are visible in~$\GP_{\uparrow\downarrow}(\omega)$.
Again, such a behavior can be attributed to the presence of the spintronic component~$D_\text{s}$ of magnetic anisotropy. Specifically, recall that unlike the intrinsic uniaxial component~$D$, which only affects the spin~$\opS_z$ of the molecular magnetic core [see Eq.~(\ref{eq:H_core})], the spintronic component~$D_\text{s}$ has the effect on the total spin of the molecule, \mbox{$\opS_z+\ops_z$}. 
Importantly, it leads to such a situation that different doublets within the \mbox{$S+1/2$} spin multiplet effectively respond in a somewhat dissimilar way to varying of~$p$. Since~\mbox{$D_\text{s}>0$}, with the increase of~$p$ the effect of intrinsic uniaxial anisotropy~(\mbox{$D<0$}) gets reduced, the system undergoes a transition from the ground state doublet~\mbox{$\ket{S_z^\text{tot}=\pm1/2}$} for small~$p$ to~\mbox{$\ket{S_z^\text{tot}=\pm5/2}$} for large~$p$. This transition, however, is not direct and it proceeds \via the doublet~\mbox{$\ket{S_z^\text{tot}=\pm3/2}$}, namely: The first (left one) of two additional features visible in Figs.~\ref{fig10}(a)-(b) appears when the doublets~\mbox{$\ket{S_z^\text{tot}=\pm1/2}$} and~\mbox{$\ket{S_z^\text{tot}=\pm3/2}$} are degenerate, and the latter doublet becomes the new ground state. With the further increase of~$p$, at some other critical value~$p$ this doublet becomes eventually degenerate with~\mbox{$\ket{S_z^\text{tot}=\pm5/2}$}, which results in formation of the right feature in Figs.~\ref{fig10}(a)-(b).
For even larger~$p$, the spintronic component~$D_\text{s}$ starts dominating over the intrinsic one~$D$ and the dynamical spin accumulation $\GP_{\uparrow\downarrow}(\omega)$ displays characteristics of the uniaxial magnetic anisotropy of the easy-axis type, compare also Fig.~\ref{fig8}(c) with Fig.~\ref{fig7}(c). Moreover, since in this limit the ground state is~\mbox{$\ket{S_z^\text{tot}=\pm5/2}$},  a significant suppression of the dynamical spin accumulation occurs.

We note that the range of spin polarization values~$\Delta p$,
for which this transition occurs, is conditioned by the interplay of the quadrupolar exchange field and the Kondo temperature.
More specifically, the Kondo effect becomes approximately suppressed
once \mbox{$D + D_\text{s} \gtrsim \TK$}.
As can be seen in Fig.~\ref{fig9}(a) for \mbox{$p<p_0$}, the Kondo temperature 
is of the order of \mbox{$\TK/\TKqd \approx 10^{-3}$}.
On the other hand, the~dependence of $D_\text{s}$
on the spin polarization can be approximated by
\mbox{$D_\text{s} \propto \Gamma^2 p^2/U$} \cite{Misiorny_NaturePhys.9/2013}.
This implies that \mbox{$\Delta p \approx \sqrt{\TK U}/\Gamma \approx 0.02$},
which agrees reasonably well with the numerical data shown in Fig.~\ref{fig10}(a).

The effect of the spintronic component of magnetic anisotropy is even more pronounced if the molecule  exhibits also the intrinsic transverse component of magnetic anisotropy~(\mbox{$E\neq0$}). The corresponding dynamical transport quantities for this case  are shown in Figs.~\ref{fig9}(e)-(h). 
First of all, the presence of non-zero $E$ results in persistence of the Kondo effect to large values of~$p$, as compared to the case of~\mbox{$E=0$} discussed above. One can see that larger values of spin polarization are needed  to suppress the Kondo peak in~$\GP(\omega)$.
One can also observe that for~\mbox{$p>p_0$} the dynamical spin accumulation~$\GP_{\uparrow\downarrow}(\omega)$ [shown in Fig.~\ref{fig9}(f)] displays a pronounced maximum for frequencies corresponding to the Kondo temperature. This resonance is a clear signature of the effective uniaxial magnetic anisotropy of the easy-axis type, as discussed in Sec.~\ref{sec:Uniaxial_and_transverse} [see especially Fig.~\ref{fig3}(c)].
Moreover, analogously to the case of~\mbox{$E=0$} [see Figs.~\ref{fig10}(a)-(b)],  $\GP_{\uparrow\downarrow}(\omega)$ is characterized by two maxima at~\mbox{$p\approx0.39$} and~\mbox{$p\approx0.44$}, which translate into a non-monotonic dependence of~$\GP(\omega)$ on the spin polarization at low frequencies~$\omega$, see Figs.~\ref{fig9}(e,g) and also the magnification of the relevant range of~$p$ in Figs.~\ref{fig10}(c)-(d). Noticeably, these features are a pure dynamical effect and they disappear in the zero-frequency limit, as illustrated by the solid line in Figs.~\ref{fig9}(g,h). 
The occurrence of these two maxima can be understood by invoking exactly the same arguments as those used above to explain the origin of the two resonances in~$\GP_{\uparrow\downarrow}(\omega)$ for \mbox{$E=0$}. 
The only difference is associated with the increased separation of the two present resonances. 
It occurs as a result of modification of states and energies of the molecule due to the presence of the transverse (second) term in Hamiltonian~(\ref{eq:H_core}), which are further renormalized non-trivially by spin-dependent electron tunneling processes.

\subsection{\label{sec:AFM-J} Antiferromagnetic coupling between the OL and the magnetic core}

%
\begin{figure}[t]
	\includegraphics[scale=1.05]{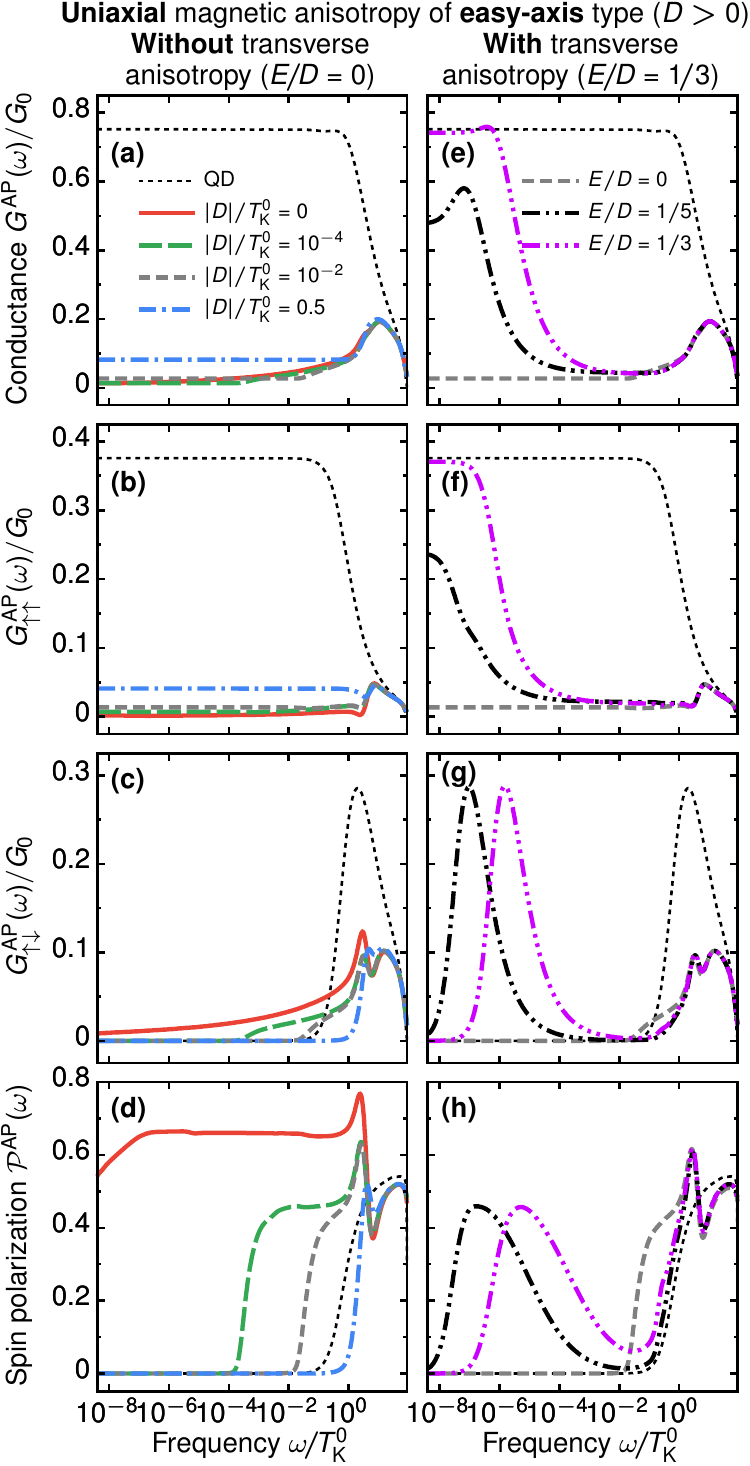}
	\caption{
	The effect of mag\-ne\-tic anisotropy on the~fre\-quency-dependent conductance of a large-spin magnetic molecule shown for the AFM $J$-coupling (\mbox{$J/\TKqd=-2.25$}) and the junction in the antiparallel (AP) magnetic configuration. 
	Note that only the uniaxial magnetic anisotropy of the easy-axis plane (\mbox{$D>0$}) is presented.
	\emph{Left column} [(a)-(d)] corresponds to the molecule exhibiting exclusively the uniaxial component of magnetic anisotropy (\mbox{$E=0$}),
	whereas \emph{right~column} [(e)-(h)] shows the effect of the transverse component for~\mbox{$D/\TKqd=10^{-2}$}.
    All other parameters are the same as~in~Fig.~\ref{fig2}.
    \label{fig11}
    }
\end{figure}

Finally, in this section we discuss the main results in the case when the coupling between the molecule's magnetic core and the orbital level is of the antiferromagnetic (AFM) type~(\mbox{$J<0$}). For this purpose, in Fig.~\ref{fig11} we show the frequency-dependent transport coefficients in the antiparallel configuration calculated for uniaxial magnetic anisotropy of the easy-axis type~(\mbox{$D>0$}) when the transverse component of magnetic anisotropy is absent~(\mbox{$E=0$}, left column) and finite~(\mbox{$E\neq0$}, right column), respectively.

We recall that a spin-isotropic molecule~($D=0$) with the AFM $J$-coupling generically exhibits the two-stage Kondo effect, so that for frequencies~$\omega$ smaller than the energy scale characteristic of the second stage of screening $\TK^*$, $\omega \lesssim \TK$, the value of conductance is expected to drop significantly, see the solid line in the left column of Fig.~\ref{fig11}. Such a suppression of conductance~$\GAP(\omega)$ [Fig.~\ref{fig11}(a)], and also of the dynamical spin accumulation~$\GAP_{\sigma\overline{\sigma}}(\omega)$ [Fig.~\ref{fig11}(c)], is a direct consequence of the AFM coupling between the spin of an electron in the orbital level and the molecule's core spin. This coupling surpasses the AFM interaction between the OL's spin and spins of conduction electrons that gives rise to the Kondo resonance. Note that for the assumed value of exchange coupling~$J$, the temperature~$\TK^*$ is relatively high, so that the conductance only slightly increases with lowering~$\omega$ due to the first-stage Kondo effect and then immediately drops down, which results in relatively low maximum around~\mbox{$\omega \approx \TK$}.

Adding the uniaxial component of magnetic anisotropy does not affect significantly the low-frequency results, as the dynamical conductance and its spin-resolved components are suppressed already for \mbox{$\omega \lesssim \TK$}, since~$\TK^*$ and~$\TK$ are in fact of a similar order. Moreover, similarly as in the case of the FM $J$-coupling, see Figs.~\ref{fig2}(a)-(d), a threshold frequency~$\omega^\ast$ arises above which the dynamical spin accumulation~$\GAP_{\sigma\overline{\sigma}}(\omega)$ is observed, as shown in Fig.~\ref{fig11}(c).
Even though the low-frequency transport is at present substantially reduced, the values of the spin polarization of the current injected into a drain electrode achieved for~\mbox{$\omega>\omega^\ast$} can be still quite large and become suppressed with increasing $D$, see Fig.~\ref{fig11}(d).

When the molecule also possesses transverse component of magnetic anisotropy, the second stage of Kondo screening can become suppressed. This is seen in Figs.~\ref{fig11}(e)-(f) where finite $E$ restores the Kondo peak at low frequencies. As expected, since now a pronounced Kondo resonance is established for \mbox{$\omega \lesssim \TK$}, the dynamical spin accumulation exhibits a maximum with its height being of the same order as in the quantum dot case, see Fig.~\ref{fig11}(g). A maximum at the same frequency is also observed in the spin polarization [Fig.~\ref{fig11}(h)].

Finally, we note that for the uniaxial magnetic anisotropy of the easy-plane type~(\mbox{$D<0$}), and also when the magnetic configuration of the device is parallel, one finds a qualitatively similar behavior to the case of ferromagnetic exchange interaction discussed in previous sections. However, due to the dominance of the second-stage Kondo effect the transport is generally suppressed.

\section{\label{sec:Con}Conclusions}

In this paper we have analyzed the dynamical transport properties
of magnetic molecules coupled to ferromagnetic electrodes in the Kondo regime.
The molecule was modeled by a LUMO level, directly coupled
to external leads and additionally coupled through a ferromagnetic exchange interaction
to the core spin of the molecule.
We have focused on the effect of dynamical spin accumulation,
which can be associated with the off-diagonal-in-spin component
of the dynamical conductance. We have in particular addressed
the question of how the dynamical spin accumulation becomes
affected by the presence of uniaxial, either of easy plane or easy axis type,
and transverse anisotropy of the molecule.
Our considerations have been performed in the linear response regime
by using the Kubo formula, while all the dynamical response functions
were determined by using the numerical renormalization group method.

We have generally shown that, in the case of antiparallel magnetic configuration of the device,
the dynamical spin accumulation 
can develop for frequencies corresponding to the energy scale responsible
for the formation of the Kondo effect, since then the spin-exchange processes are most effective.
A local maximum in $G_{\sigma\overline{\sigma}}(\omega)$ thus develops
for resonant frequency~$\omegar$, which is of the order of $\TK$.
The width of this local maximum depends in turn 
on the energy scale where the Kondo state is formed.
Consequently, while for spin isotropic molecules
dynamical spin accumulation exhibits a broad maximum,
in the presence of magnetic anisotropy the width of this maximum becomes reduced.
Another important energy scale describing the behavior
of the dynamical spin accumulation denoted by $\omega^*$
corresponds to the frequency below which the conductance reaches a plateau.
Then, the spin-flip processes become quenched,
such that $G_{\sigma\overline{\sigma}}(\omega)$ vanishes.
On the other hand, the height of the maximum in dynamical spin accumulation
turned out to depend strongly on the magnitude of the Kondo effect.
We have shown that if the Kondo resonance develops fully
for \mbox{$\omega < \TK$}, the spin accumulation exhibits
a local maximum, the height of which is of the same order as in the 
case of quantum dots. However, when due to the presence of magnetic anisotropy
the Kondo effect cannot fully emerge, the effect of dynamical spin accumulation
correspondingly becomes reduced.

When the junction is in the parallel magnetic configuration,
the situation drastically changes, since then the spin-resolved renormalization
of the molecule comes into play. As a result, dynamical transport properties,
and also the dynamical spin accumulation,
are conditioned by the interplay of the quadrupolar exchange field
and Kondo correlations. For the out-of-plane type of magnetic anisotropy we have predicted
a large dynamical spin accumulation if the transverse anisotropy is present.
On the other hand, for the easy plane type of magnetic anisotropy, we have shown that, 
with increasing the spin polarization of the leads, there is a transition
from a spin doublet to higher-spin state, which results in a suppression
of the Kondo resonance. Interestingly, the suppression 
of the low-frequency conductance occurs in a non-monotonic fashion,
which is an indication of a highly nontrivial competition between the intrinsic
and spintronic contributions to the magnetic anisotropy of the molecule.

In addition, we have also performed the analysis
assuming that the exchange coupling between the orbital level
and magnetic core is of antiferromagnetic type. Then,
in the absence of magnetic anisotropy and for nonmagnetic leads,
the system exhibits two-stage Kondo effect.
We have shown that finite magnetic anisotropy of the molecule
leads to the suppression of both the frequency-dependent conductance
and the effect of dynamical spin accumulation.

Finally, we recall that all the considerations presented in this paper
were performed for the particle-hole symmetry point of the model.
However, it is worth emphasizing that the results discussed
in the case of the antiparallel magnetic configuration are qualitatively valid
also out of the particle-hole symmetry point, provided the system
is in the Kondo regime (the orbital level is singly occupied).
On the contrary, in the case of the parallel magnetic configuration,
detuning from the symmetry point results in a development
of the effective dipolar exchange field. Because this field acts in a similar fashion to 
external magnetic field and splits the orbital level, 
most of the presented effects will become suppressed
once the magnitude of exchange field surpasses
the relevant energy scales for considered phenomena.
Last but not least, we want to notice that, since the position of the orbital
level can be tuned by a gate voltage, the effects predicted in this paper
can be tested in near future with present-day experimental techniques.

\acknowledgments

Work supported by the Polish Ministry of Science and Education
as Iuventus Plus project (IP2014 030973) in years 2015-2017 (A.P., M.M.)
and the Polish National Science Centre from funds awarded
through the decision No. DEC-2013/10/E/ST3/00213 (I.W.).
M.M. also acknowledges financial support from the Polish Ministry of Science
and Education through a young scientist fellowship (0066/E- 336/9/2014)
and from the Knut and Alice Wallenberg Foundation.
Part of calculations was performed at Pozna\'n Supercomputing and Networking Center.



%

\end{document}